\documentclass[journal,twoside]{IEEEtran}
\IEEEoverridecommandlockouts
\usepackage{balance}
\usepackage{amsmath}
\usepackage{amsthm}
\usepackage[numbers,sort&compress]{natbib}
\usepackage{cases}
\usepackage{amssymb}
\usepackage{algorithm}
\usepackage{algorithmic}
\usepackage{graphicx}
\usepackage{subfigure} 
\usepackage{color}
\usepackage{enumerate}

\newtheorem{thm}{Theorem}
\newtheorem{lem}{Lemma}
\ifCLASSINFOpdf
\else
\fi
\DeclareMathOperator{\diag}{diag}
\DeclareMathOperator{\sign}{sign}
\DeclareMathOperator{\Tr}{Tr}
\DeclareMathOperator{\rank}{rank}

\begin{document}
%
\title{Gridless Parameter Estimation for One-Bit MIMO Radar with Time-Varying Thresholds}
%
%
%

\author{Feng~Xi,~\IEEEmembership{Member,~IEEE,}
        Yijian~Xiang,~\IEEEmembership{Student Member,~IEEE,}
        Shengyao Chen,~\IEEEmembership{Member,~IEEE,}
        and~Arye~Nehorai,~\IEEEmembership{Life~Fellow,~IEEE}
\thanks{F. Xi and S. Chen are with the Department
of Electronic Engineering, Nanjing University of Science and Technology, Nanjing 210094 China (email:xifeng@njust.edu.cn).}
\thanks{Y. Xiang and A. Nehorai are with the Department of Electrical and Systems Engineering, Washington University in St. Louis, St. Louis, MO 63130 USA.}
\thanks{The work of F. Xi was supported by the China Scholarship Council for his visit at Washington University in St. Louis and supported in part by National Science Foundation of China (No. 61571228). }
}

\maketitle

\begin{abstract}
We investigate the one-bit MIMO (1b-MIMO) radar that performs one-bit sampling with a time-varying threshold in the temporal domain and employs compressive sensing in the spatial and Doppler domains.
The goals are to significantly reduce the hardware cost, energy consumption, and amount of stored data.
The joint angle and Doppler frequency estimations from noisy one-bit data are studied.
By showing that the effect of noise on one-bit sampling is equivalent to that of sparse impulsive perturbations,
we formulate the one-bit $\ell_1$-regularized atomic-norm minimization (1b-ANM-L1) problem to achieve gridless parameter estimation with high accuracy.
We also develop an iterative method for solving the 1b-ANM-L1 problem via the alternating direction method of multipliers.
The Cram$\acute{\text{e}}$r-Rao bound (CRB) of the 1b-MIMO radar is analyzed, and the analytical performance of one-bit sampling with two different threshold strategies is discussed.
Numerical experiments are presented to show that the 1b-MIMO radar can achieve high-resolution parameter estimation with a largely reduced amount of data.
\end{abstract}


%
\IEEEpeerreviewmaketitle

\section{Introduction}
\IEEEPARstart{M}{ultiple} input multiple output (MIMO) \cite{Fishler:MIMORadar-04,Li:MIMORadar-07,Haimovich:MIMORadar-08} radar, which employs multiple antenna elements at both the transmitter and receiver and radiates a set of mutually orthogonal waveforms, has attracted significant attention in recent years due to its superior resolution performance.
Depending on the deployment of their multiple antennas, two typical MIMO radar architectures are considered, including collocated MIMO radar \cite{Li:MIMORadar-07}, with collocated transmit and receive antennas, and distributed MIMO radar \cite{Haimovich:MIMORadar-08}, with widely separated antennas.
In this paper, we mainly focus on collocated MIMO radar.

In collocated MIMO radar \cite{Li:MIMORadar-07}, the waveform diversity is exploited by employing a bank of matched filters at each receive antenna
to separate different waveforms and thus generate a large virtual array.
Then all the data collected at each antenna is transmitted to a fusion center for sophisticated processing.
As a result, MIMO radar achieves superior for flexibility and resolution performance.
However, the complexity of such a MIMO radar system and its corresponding signal processing is inevitably greater than that of a conventional radar system.
The main challenges of MIMO radar are coping with the cost, energy consumption, data transmission volume, and computational load of complicated systems. 

Inspired by the recent development of compressive sensing (CS), a variety of sub-Nyquist sampling structures \cite{Mishali:Xampling-11,Becker:RMPI-11,Xi:QuadCS-14} with their corresponding signal processing methods \cite{Eldar:DoppFoc-14,Yoo2012-RMPI,Liu:QuadCS-15} 
have been designed for radar systems to significantly reduce the sampling rate while maintaining the performance.
In the realm of MIMO radars, CS is applied to either compress the amount of sampled data at each antenna \cite{Yu2010-JSTSP,Sun:MIMO-MC-15} or to reduce the number of antennas used in a transmitter and receiver \cite{Rossi2014-TSP}.
These applications are respectively equivalent to implementing CS in either the temporal domain or the spatial domain.
In \cite{Cohen2018-TSP}, a sub-Nyquist MIMO radar (SUMMeR) system that performs both temporal and spatial compression was proposed to further simplify a conventional MIMO radar system.
In \cite{Cohen2016-ReducedTime}, the reduced time-on-target technique was introduced to perform compression in the Doppler domain.
A recent work \cite{Na-SPL2019} extended the SUMMeR system to perform the temporal, spatial, and Doppler compression simultaneously.
Interested readers are referred to a survey paper \cite{Cohen2018-SubNyquistSPM} for an overall perspective on different compression techniques used in radar systems.
However, all these works are developed under the assumption of infinite precision sampling, and they do not address the effect of finite bit quantization. 
As a matter of fact, sampling and quantization are indispensable in signal acquisition.

As we know, the energy consumption and cost of an analog-to-digital converter (ADC) grow exponentially with the bit depth \cite{Walden99-ADCJSAC,Rondeau2005-ADCSPM}.
Therefore, low-bit quantization may be remarkably useful in digital systems, due to its low cost, low energy consumption, and low data volume.
As an extreme case of low-bit quantization, one-bit sampling, which simplifies the conventional ADC to a simple comparator, has attracted much research interest recently.
Several theoretical works \cite{Boufounos:1bit-2008,Laska:Trust-TSP-2011,
Jacques:1bit-TIT-2013} have proved that it is indeed possible to recover the signal from its one-bit measurements by exploiting the signal sparsity.
Due to its low-complexity hardware and low energy consumption, one-bit sampling has many applications, including spectrum sensing \cite{Gianelli-CAMSAP2017,Qian:ADMM-ICASSP-2017,Fu2018-TSP},
DOA estimation \cite{Bar-Shalom-TAES2002,Yu:DOA-SPL-2016}, pulse-Doppler radar processing \cite{Li:Asilomar-2016,Zahabi:ICASSP-2017,Xi2018-SAM}, massive MIMO channel estimation \cite{Li2017-OnebitMassiveMIMO-TSP,Stein2018-1bit-TSP,Wang2018-TVT}, to name a few.
It is worth noting that one-bit sampling is accomplished via comparing the signal to some reference level and outputting 1-bit data to denote whether the signal is above or below the reference level.
In most of the previous works, the commonly used reference level is zero.
Unfortunately, this reference level is unable to recover the signal amplitude due to the loss of amplitude information \cite{Boufounos:1bit-2008,Knudson2016-TIT}.
To enable the accurate amplitude estimation, one-bit sampling with a dithered reference level must be used \cite{Cambareri2017,Jacques2017-TIT,Xu-2018arXiv}, which is equivalent to comparing the signal to a set of time-varying thresholds \cite{Gianelli-CAMSAP2017,Li:Asilomar-2016,Zahabi:ICASSP-2017,Khobahi:ACSSC2018}.
In radar applications, it is especially important to recover the amplitude information, since the amplitudes of radar echoes are critical in detecting and classifying the targets.

Despite the above benefits, one of the main challenges in one-bit sampling is developing efficient methods to recover the original signal or extract desired information from the one-bit sampled data.
Following the idea of CS, a large variety of sparsity-based methods for 1-bit signal recovery have been proposed \cite{Laska:Trust-TSP-2011,Jacques:1bit-TIT-2013,Plan-TIT2013,Boufounos-2009Conf,Yan-TSP2012}, which attempt to find the sparsest signal, 
for which the one-bit measurements are consistent with the measurements of the recovered signal.
However, due to the discretization of the parameter space in sparse recovery, these sparsity-based methods suffer from the off-grid problem \cite{Chi:BasisMismatch-11}.
To avoid this problem, another commonly-used method maximizes the likelihood function \cite{Wang2018-TVT,Gianelli-CAMSAP2017,Ren:MM-RELAX-TSP2019}.  
But the exhaustive search to find the peaks of the likelihood function leads to a high computational burden, especially when it is applied to multi-dimensional parameter estimation.

Recently, atomic norm-based super-resolution theory \cite{Candes:SuperRes-14,Tang:offgrid-13,Bhaskar-TSP2013,Yang:Gridless-15,Yang:MultiToep-TSP2016,Zhang:D-ANM-SP2019} has emerged as an effective approach to allow gridless sparse recovery.
Inspired by this approach, an atomic norm soft thresholding algorithm \cite{Fu2018-TSP} was proposed to recover spectrally-sparse signals from their 1-bit measurements,
and \cite{Xi2018-SAM} extended the work of \cite{Fu2018-TSP} to two-dimensional parameter estimation.
The success of these works lies in designing a surrogate signal formed by one-bit measurements to approximate the original signal.
However, these works concentrate on one-bit sampling with a zero threshold, and it is still unclear how to design an effective surrogate signal for one-bit sampling with time-varying thresholds. 

In this paper, we seek to understand the effect of one-bit sampling in MIMO radar, 
as well as to develop an efficient method to jointly estimate the angle and Doppler frequency from noisy one-bit sampled data.
In particular, we consider a MIMO radar system which employs one-bit sampling with time-varying thresholds in the temporal domain and performs CS in the spatial and Doppler domains, referred to as 1b-MIMO radar.
To mitigate the off-grid problem and achieve high-resolution parameter estimation, a one-bit $\ell_1$-regularized atomic-norm minimization (1b-ANM-L1) formulation is proposed.
We also develop an iterative algorithm via the alternating direction method of multipliers (ADMM) \cite{Boyd:ADMM-11} to
obtain a computationally efficient solution to the 1b-ANM-L1 problem.
Note that, although the ADMM method has been applied to $\ell_1$-regularized atomic-norm minimization in \cite{Zheng-TSP2017}, it is still unclear how the ADMM method can be generalized to the one-bit case.
In addition, to understand the effect of one-bit sampling on parameter estimation, we establish the Cram$\acute{\text{e}}$r-Rao bound (CRB) for the 1b-MIMO radar.

Our main contributions can be summarized as follows:
\begin{enumerate}
\item We present the 1b-ANM-L1 method to jointly estimate the angle and Doppler frequency parameters from  noisy one-bit data.
The proposed method is founded on the fact that the effect of noise on one-bit sampling is equivalent to that of sparse impulsive perturbations.

\item A computationally efficient algorithm to solve the 1b-ANM-L1 problem is developed based on the ADMM method, in which the closed-form computation in each iteration is explicitly derived. 

\item The CRB analysis for the 1b-MIMO radar is established, in which we show that the Fisher information matrix for the 1b-MIMO radar is a weighted version of that for the unquantized MIMO radar. 
Based on the CRB analysis, we discuss the effects of two different threshold strategies, the random uniform threshold (RUT) and the random Gaussian threshold (RGT), on parameter estimation.

\item Numerical simulations are provided to demonstrate the performance of the 1b-MIMO radar. We also compare the performance of the 1b-MIMO radar with that of its high-bit quantized rivals. 
The results show that it is indeed possible to achieve high-resolution angle and Doppler frequency estimation while largely reducing the amount of data.

\end{enumerate}

\emph{Notations}:
We use lower-case (upper-case) bold characters to denote vectors (matrices).
In particular, $\mathbf{I}_N$ denotes the $N \times N$ identity matrix.
$\mathbb{R}$ and $\mathbb{C}$ denote the sets of real and complex numbers, respectively.
$(\cdot)^{\ast}$, $(\cdot)^T$ and $(\cdot)^H$ denote the complex conjugate, matrix transposition, and Hermitian transposition, respectively.
$\lceil\cdot\rceil$ and $\lfloor\cdot\rfloor$  respectively denote the ceiling and the floor functions.
For a vector $\mathbf{x}$, $[\mathbf{x}]_n$ denotes the $n$-th element of $\mathbf{x}$,
and $\textrm{diag}(\mathbf{x})$ represents a diagonal matrix with $\mathbf{x}$ as its diagonal elements.
For a matrix $\mathbf{X}$, $\textrm{vec}(\mathbf{X})$ denotes the vectorization operator that turns the matrix $\mathbf{X}$ into a vector by stacking all columns on top of the another.
$\textrm{Tr}(\cdot)$ denotes the matrix trace.
$\otimes$ and $\odot$ represent the Kronecker product and Hadamard product, respectively.
For two matrices $\mathbf{A}$ and $\mathbf{B}$, $\langle\mathbf{A},\mathbf{B}\rangle=\Tr(\mathbf{B}^H\mathbf{A})$.
For positive semidefinite matrices $\mathbf{A}$ and $\mathbf{B}$, $\mathbf{A}\succeq\mathbf{B}$ means $\mathbf{A}-\mathbf{B}$ is positive semidefinite.

\section{Classic MIMO Radar}

Colocated MIMO radar consists of two uniform linear arrays (ULAs) with $N$ receive antennas, spaced by $d_r = \frac{\lambda}{2}$, and $M$ transmit antennas, spaced by $d_t = N\frac{\lambda}{2}$.
Here, $\lambda$ is the wavelength of the carrier signal.
A set of $M$ narrow-band and orthogonal waveforms, denoted as $s_1(t),\cdots, s_M(t)$, are transmitted in pulses, with a pulse repetition interval (PRI) $T_{\text{PRI}}$.
We assume that each coherent processing interval (CPI) includes $Q$ pulses, i.e., $T_{\text{CPI}}=QT_{\text{PRI}}$.

Now suppose that there are $K$ non-fluctuating point targets satisfying the stop-and-hop assumption \cite{Richards:RSP-05} in the far field at angles $\theta_k$, $k=1,\cdots,K$, each moving with speed $\nu_k$.
All the targets are assumed to fall in the same range bin.
To simplify the expression, we define $\vartheta_k\triangleq d_r\sin(\theta_k)/\lambda$ and $\upsilon_k\triangleq 2\nu_kT/\lambda$ as the normalized spatial frequency and normalized Doppler frequency, respectively.
Then the received waveforms collected at the $N$ receive antennas during the $q$-th pulse can be formulated as
\begin{equation}
\label{eqn_yqt}
\mathbf{y}_q(t) =\sum_{k=1}^K\beta_ke^{j{2\pi}(q-1)\upsilon_k}\mathbf{b}(\theta_k)\mathbf{a}^T(\theta_k)\mathbf{s}(t)+\mathbf{w}_q(t),
\end{equation}
where $\beta_k$ is the reflection coefficient of the $k$-th target, $\mathbf{s}(t)=[s_1(t),\cdots,s_M(t)]^T$ is the transmitting signal vector, and $\mathbf{a}(\theta) = [1,e^{j{2\pi}N\vartheta},\cdots,e^{j{2\pi}N(M-1)\vartheta}]^T$ and
$\mathbf{b}(\theta)=[1,e^{j{2\pi}\vartheta},\cdots,e^{j{2\pi}(N-1)\vartheta}]^T$ are the transmit and receive steering matrices, respectively.
$\mathbf{w}_q(t)\in\mathbb{C}^{N}$ denotes the noise vector received by the $N$ receive antennas during the $q$-th pulse.

After receiving the waveforms, each receive antenna uses a high-bit ADC to sample and quantize the received waveforms, obtaining $L$-length sampled data during each PRI, where $L=\lfloor T_{\text{PRI}}/T_s\rfloor+1$, with $T_s$ being the sampling interval.
The effect of high-bit quantization is simply modeled as an additive quantization error.
Therefore the received data at the $N$ receive antennas during the $q$-th pulse can form the following data matrix: 
\begin{equation}
\label{eqn_Yq}
\mathbf{Y}_q =\mathbf{B}\mathbf{\Sigma}\mathbf{\Delta}_q\mathbf{A}^T\mathbf{S}+\mathbf{W}_q,
\end{equation}
where $\mathbf{A}\triangleq [\mathbf{a}(\theta_1),\cdots,\mathbf{a}(\theta_K)]\in\mathbb{C}^{M\times K}$,
$\mathbf{B}\triangleq [\mathbf{b}(\theta_1),\cdots,\mathbf{b}(\theta_K)]\in\mathbb{C}^{N\times K}$, and
$\mathbf{\Delta}_q$ and $\mathbf{\Sigma}$ are two diagonal matrices given as
\begin{equation*}
\mathbf{\Delta}_q=\textrm{diag}([e^{j{2\pi}(q-1)\upsilon_1},\cdots,e^{j{2\pi}(q-1)\upsilon_K}]),
\end{equation*}
\begin{equation*}
\mathbf{\Sigma}=\textrm{diag}([\beta_1,\cdots,\beta_K]).
\end{equation*}
$\mathbf{S}=[\mathbf{s}(0),\mathbf{s}(T_s),\cdots,\mathbf{s}((L-1)T_s)]\in\mathbb{C}^{M\times L}$ is the sampled data of the transmitting waveforms.
$\mathbf{W}_q \in \mathbb{C}^{N\times L}$ is the matrix accounting for the effect of noise and quantization error.

Due to the orthogonality of the transmitting waveforms, i.e., $\mathbf{S}\mathbf{S}^H=\mathbf{I}_M$,
each receive antenna can employ a bank of $M$ matched filters to separate the information from the $M$ transmit antennas. 
As a result, $N$ receive antennas can obtain a total of $MN$ channels, which forms a virtual ULA with length $MN\lambda/2$. 

After the matched filtering, the received data $\mathbf{Y}_q$ becomes
\begin{equation}
\label{eqn_YqMF}
\begin{split}
\mathbf{Y}^{\textrm{MF}}_q&=\mathbf{Y}_q\mathbf{S}^H\\
&=\mathbf{B}\mathbf{\Sigma}\mathbf{\Delta}_q\mathbf{A}^T+\mathbf{W}^{\textrm{MF}}_q,
\end{split}
\end{equation}
where $\mathbf{W}^{\textrm{MF}}_q=\mathbf{W}_q\mathbf{S}^H$. 
Then the set of matrices $\{\mathbf{Y}^{\textrm{MF}}_q\}_{q=1}^{Q}$ are forwarded to the fusion center for subsequent processing.

By stacking the matrix $\mathbf{Y}^{\textrm{MF}}_q$ into vector $\mathbf{y}^{\textrm{MF}}_q$, the $Q$ pulses yield the following $MN\times Q$ matrix
\begin{equation}
\label{eqn_Y}
[\mathbf{y}^{\textrm{MF}}_1,\cdots,\mathbf{y}^{\textrm{MF}}_Q]=\mathbf{C}\mathbf{\Sigma}\mathbf{D}^H
+[\mathbf{w}^{\textrm{MF}}_1,\cdots,\mathbf{w}^{\textrm{MF}}_Q],
\end{equation}
where $\mathbf{C}=[\mathbf{c}(\theta_1),\cdots,\mathbf{c}(\theta_K)]$ with $\mathbf{c}(\theta)=\mathbf{a}(\theta)\otimes\mathbf{b}(\theta)$, 
$\mathbf{D}=[\boldsymbol{\delta}_1,\cdots,\boldsymbol{\delta}_Q]^H \in \mathbb{C}^{Q\times K}$ with $\boldsymbol{\delta}_q=[e^{j{2\pi}(q-1)\upsilon_1},\cdots,e^{j{2\pi}(q-1)\upsilon_K}]^T$, and $\mathbf{w}^{\textrm{MF}}_q=\textrm{vec}(\mathbf{W}^{\textrm{MF}}_q)$.
Then the joint angle and Doppler frequency estimation problem can be equivalent to estimating the two-dimensional frequencies \cite{Kay-TASSP1990, Hua-TSP1992}.

In classic MIMO radars, to achieve high angular resolution, the array aperture of the virtual ULA has to be large, which inevitably increases the complexity of the MIMO radar system.
Similarly, to achieve high Doppler resolution, a large number of pulses must be sent out, leading to a longer CPI and producing huge amounts of data.
In a hostile environment, a longer CPI can also increase the risk of interception by opponents. 
The aim of this paper is to simplify the systematic complexity and reduce the volume of data in the MIMO radar, while preserving high resolution performance.

\section{One-Bit MIMO Radar}
In this section, we propose a 1b-MIMO radar system that employs one-bit ADC to simplify the system and reduce the amount of data.
The sparse antenna array (SAA) technique and reduced time-on-target (RTT) technique are also considered to perform CS in the spatial and Doppler domains, respectively.

Specifically, the MIMO radar applies the SAA technique to randomly select $T<M$ transmit antennas and $R<N$ receive antennas during each PRI.
The antenna selection function can be implemented by using a set of low-complexity radio frequency (RF) switches \cite{Wang2014-TSP}.
Therefore, two index sets, $\Omega_q^t\subset\{1,2,\cdots,M\}$ and $\Omega_q^r\subset\{1,2,\cdots,N\}$, corresponding to the transmit and receive antennas, respectively, are chosen during the $q$-th pulse. 

Let $\mathbf{\Gamma}_{q}^t\in \{0,1\}^{T\times M}$ be a selection matrix for the transmit array, which consists of the rows of $\mathbf{I}_{M}$ indexed by the set $\Omega_q^t$.
Similarly, let $\mathbf{\Gamma}_{q}^r\in \{0,1\}^{R\times N}$ be a selection matrix for the receive array.
Then the transmit steering matrix and the receive steering matrix become $\mathbf{\tilde{A}}_{q}=\mathbf{\Gamma}_{q}^t\mathbf{A}$ and $\mathbf{\tilde{B}}_{q}=\mathbf{\Gamma}_{q}^r\mathbf{B}$, respectively.
The received data at the $R$ receive antennas during the the $q$-th pulse, denoted as $\mathbf{\tilde{Y}}_q\in\mathbb{C}^{R\times L}$, can be represented as
\begin{equation}
\label{eqn_YSubq}
\mathbf{\tilde{Y}}_q
=\mathbf{\tilde{B}}_{q}\mathbf{\Sigma}\mathbf{\Delta}_q\mathbf{\tilde{A}}^T_{q}\mathbf{\tilde{S}}_q
+\mathbf{\tilde{W}}_q,
\end{equation}
where $\mathbf{\tilde{S}}_{q}=\mathbf{\Gamma}_{q}^t\mathbf{S}$
and 
$\mathbf{\tilde{W}}_q=\mathbf{\Gamma}_{q}^r\mathbf{W}_q$ denote the transmitting signals at the $T$ transmit antennas and the noise at the $R$ receive antennas, respectively.

To reduce the time on target, the MIMO radar randomly selects a subset of the $Q$ PRIs at which to send out pulses.
Let $\Omega^p\subset\{1,2,\cdots,Q\}$ with $|\Omega^p|=P<Q$ be the set of pulses sent out by the transmitter.
In this case, the work period of the MIMO radar is reduced from $QT_{\text{PRI}}$ to $PT_{\text{PRI}}$, which significantly cuts down the energy consumption and lowers the probability of interception in a hostile environment.
During the period of $(Q-P)T_{PRI}$ when no pulse is sent out, the radar can work in other modes, which enables multi-function radar \cite{Li-TAES2017,Qian-TSP2018}.
With the RTT technique, the received data will be $\{\mathbf{\tilde{Y}}_q\}_{q\in\Omega^p}$.
In this paper, to keep the same CPI, we assume that the set $\Omega^p$ will always include the elements 1 and $Q$, i.e., the transmitter always sends out pulses during the first and the last PRIs in one CPI.

In one-bit sampling, a pair of one-bit ADCs is used to sample and quantize the real and imaginary parts of received signals separately.
Let $\mathcal{Q}_1(\cdot)=\sign(\Re\{\cdot\})+j\sign(\Im\{\cdot\})$ be the complex one-bit quantization operator, 
where $\sign(\cdot)$ denotes the sign function applied element-wise to any vector or matrix.
Then the one-bit data matrix at the $R$ receive antennas during the the $q$-th pulse, denoted as $\mathbf{Z}_q$, is given as
\begin{equation}
\label{eqn_Zq}
{\mathbf{Z}}_{q}= \mathcal{Q}_1(\mathbf{\tilde{Y}}_q-\mathbf{H}_q),
\end{equation}
where $\mathbf{H}_q\in\mathbb{C}^{R\times L}$ represents the known threshold to which the one-bit quantization applies.
If $\mathbf{H}_q$ is zero, then an identical zero threshold is applied.
We will discuss threshold strategies in Section VI.

In Table I, we summarize the different temporal/Doppler/spatial sampling strategies used in classic MIMO radar and the 1b-MIMO radar.
With these sampling strategies, the system complexity as well as the amount of data forwarded to the fusion center is largely reduced.
However, due to the one-bit sampling, the classic matched filtering-based processing does not work here. 
One of the main challenges for the 1b-MIMO radar is to detect and resolve the the set of parameters $\{(\theta _k,\nu _k)\}_{k = 1}^K$ from the extremely limited one-bit data $\{\mathbf{Z}_q\}_{q\in\Omega^p}$.  

\begin{table}[]
\caption{The different sampling strategies between classic MIMO radar and one-bit MIMO radar.}
\centering
\renewcommand\arraystretch{1.25}
\begin{tabular}{|l|l|l|l|}
\hline
 Domain             & Classic MIMO Radar              & 1b-MIMO Radar      \\ \hline
Temporal            & Nyquist/High-bit sampling       & Nyquist/1-bit sampling      \\ \hline
Doppler             & Uniform PRI                     & RTT-based CS   \\ \hline
Spatial             & Uniform linear array            & SAA-based CS   \\ \hline
\end{tabular}
\end{table}

\section{Joint Angle and Doppler Frequency Estimation via Atomic-Norm Minimization}
In this section, we formulate the joint angle and Doppler frequency estimation problem as an atomic norm minimization problem.
To deal with the noise in the one-bit data, we show that the effect of noise in one-bit data can be replaced by impulsive perturbations.
As a result, an $\ell_1$-regularized atomic norm minimization method is proposed to recover the target information and the impulsive perturbation simultaneously from the noisy one-bit data.

\vspace{-0.3cm}
\subsection{Atomic-Norm Formulation}\label{Sec:ANM_A}

By vectorizing the unquantized data matrix $\mathbf{\tilde{Y}}_q$, we can rewrite (\ref{eqn_YSubq}) in the following form:
\begin{equation}
\label{eqn_ysubq}
\mathbf{\tilde{y}}_q=\mathbf{M}_q
\mathbf{C}\mathbf{\Sigma}\mathbf{D}^H\mathbf{e}_q
+\mathbf{\tilde{w}}_q,
\end{equation}
where $\mathbf{M}_q=\mathbf{\tilde{S}}^T_q\mathbf{\Gamma}_{q}^t\otimes \mathbf{\Gamma}_{q}^r\in\mathbb{C}^{LR\times MN}$,
$\mathbf{e}_q$ denotes the $q$-th column of the $Q\times Q$ identity matrix $\mathbf{I}_Q$,
and $\mathbf{\tilde{w}}_q$ is the vectorized version of $\mathbf{\tilde{W}}_q$.
To simplify the representation, we define the matrix $\mathbf{X}\triangleq\mathbf{C}\mathbf{\Sigma}\mathbf{D}^H\in\mathbb{C}^{MN\times Q}$ and the operator $\mathcal{F}_q(\mathbf{X})\triangleq\mathbf{M}_q\mathbf{X}\mathbf{e}_q$, i.e.,
\begin{equation}
\label{eqn_ysubq2}
\mathbf{\tilde{y}}_q=\mathcal{F}_q(\mathbf{X})+\mathbf{\tilde{w}}_q,
\end{equation}

Then the vectorized version of the one-bit data matrix $\mathbf{Z}_q$, denoted as $\mathbf{z}_q$, can be expressed as
\begin{equation}
\label{eqn_Qyq}
\mathbf{z}_q=\mathcal{Q}_1(\mathbf{\tilde{y}}_q-\mathbf{h}_q),
\end{equation}
where $\mathbf{h}_q$ is the vectorized version of $\mathbf{H}_q$.

From (\ref{eqn_ysubq}), it is seen that the angle and Doppler frequency parameters of the targets are completely determined by the matrix $\mathbf{X}$.
In fact, the matrix $\mathbf{X}$ is equivalent to the noise-free data matrix in (\ref{eqn_Y}).
By considering the structures of $\mathbf{C}$, $\mathbf{\Sigma}$, and $\mathbf{D}$, we can derive
\begin{equation}
\label{eqn_X}
\mathbf{X} =\sum_{k=1}^K \beta_k\mathbf{c}(\theta_k)\mathbf{d}^H(\nu_k),
\end{equation}
where $\mathbf{d}(\nu_k) = [1,e^{j2\pi\upsilon_k},\cdots,e^{j2\pi(Q-1)\upsilon_k}]^H$.
It is shown that the rank of the matrix $\mathbf{X}$ is no larger than $K$.
In sparse target scenarios, i.e., $K\ll \min\{MN,Q\}$, $\mathbf{X}$ is a low-rank matrix.
Therefore, our problem now is equivalent to recovering the low-rank matrix $\mathbf{X}$ from a set of one-bit measurements $\{\mathbf{z}_q\}_{q\in\Omega^p}$.
In what follows, we define the atomic norm to enforce the structure constraint of $\mathbf{X}$.

According to \cite{Xi:ICASSP-2017,Tian-ICASSP2017}, we can define a set of atoms to describe the structure of $\mathbf{X}$:
 \begin{equation*}
  \mathcal{A}\triangleq \{\mathbf{A}(\boldsymbol{\varphi},\phi)=e^{j\phi}{\mathbf{w}(\varphi_1)}{\mathbf{v}^H(\varphi_2)}:\varphi_1,\varphi_2 \in \mathbb{T}, \phi \in \mathbb{S}\},
  \end{equation*}
where $\boldsymbol{\varphi}=\{\varphi_1,\varphi_2\}$, 
$\mathbb{T}\triangleq (0,1]$, 
$\mathbb{S}\triangleq(0,2\pi]$, 
$\mathbf{w}(\varphi)=[1,e^{j2\pi\varphi},\cdots,e^{j2\pi(MN-1)\varphi}]^T\in\mathbb{C}^{MN}$, 
and $\mathbf{v}(\varphi)=[1,e^{j2\pi\varphi},\cdots,e^{j2\pi(Q-1)\varphi}]^T\in\mathbb{C}^{Q}$. 
Then the atomic $l_0$ norm of the matrix $\mathbf{X}$ is defined as the smallest number of atoms in $\mathcal{A}$ that can express $\mathbf{X}$:
\begin{equation}
   \label{eqn_Atoml0norm}
   \|\mathbf{X}\|_{\mathcal{A},0}=\inf_{\mathbf{A}(\boldsymbol{\varphi}_k,\phi_k)\in\mathcal{A}} \bigg\{\mathcal{K}:\mathbf{X}=\sum_{k=1}^{\mathcal{K}}{a_{k}\mathbf{A}(\boldsymbol{\varphi}_k,\phi_k)},a_k>0\bigg\}.
  \end{equation}
As shown in  \cite{Xi:ICASSP-2017},
$\|\mathbf{X}\|_{\mathcal{A},0}$ can be cast as an equivalent rank minimization problem,
\begin{equation}
  \label{eqn_SDP}
  \begin{split}
\|\mathbf{X}\|_{\mathcal{A},0}=&\min_{\mathbf{u}_1,\mathbf{u}_2}\bigg\{\rank(\mathbf{H})\bigg|\\
&\mathbf{H}=\begin{bmatrix}
\mathcal{T}(\mathbf{u}_1) & \mathbf{X}\\
\mathbf{X}^H & \mathcal{T}(\mathbf{u}_2)
\end{bmatrix}\succeq0\bigg\},
\end{split}
  \end{equation}
where $\mathcal{T}(\mathbf{u})$ denotes a Toeplitz matrix with $\mathbf{u}^T$ as its first row.
For the sake of completeness, we provide the proof of (\ref{eqn_SDP}) in Appendix A.

If the unquantized data $\{\tilde{\mathbf{y}}_q\}_{q\in\Omega^p}$ is considered, we can formulate the low-rank matrix recovery problem as
\begin{equation}
\label{opt_MinANMunq}
\begin{split}
\min_{\mathbf{u}_1,\mathbf{u}_2,\mathbf{X}}& \Tr(\mathcal{T}(\mathbf{u}_1))+\Tr(\mathcal{T}(\mathbf{u}_2)),\\
\text{s.t.}\quad&
\begin{bmatrix}
\mathcal{T}(\mathbf{u}_1) & \mathbf{X}\\
\mathbf{X}^H & \mathcal{T}(\mathbf{u}_2)
\end{bmatrix}\succeq0\\
& \|\tilde{\mathbf{y}}_q-\mathcal{F}_q(\mathbf{X})\|_2\leq\epsilon, \text{for all}\; q\in\Omega^p,
\end{split}
\end{equation}
where the non-convex rank minimization is relaxed to trace minimization, and $\epsilon$ is an algorithmic parameter determined by the noise.

For the one-bit data $\{\mathbf{z}_q\}_{q\in\Omega^p}$, one of the main challenges is how to enforce the constraint between the one-bit data and the recovered unquantized data.
Let us define the notation $\mathbf{a}\in\mathcal{S}(\mathbf{b})$ representing that $\Re\{\mathbf{a}\}\odot\Re\{\mathbf{b}\}\geq 0$ and 
$\Im\{\mathbf{a}\}\odot\Im\{\mathbf{b}\}\geq 0$.
Then, in the noise-free case,
recovering the low-rank matrix $\mathbf{X}$ from one-bit data can be formulated as
\begin{equation}
\label{opt_MinANM}
\begin{split}
\min_{\mathbf{u}_1,\mathbf{u}_2,\mathbf{X}}& \Tr(\mathcal{T}(\mathbf{u}_1))+\Tr(\mathcal{T}(\mathbf{u}_2)),\\
\text{s.t.}\quad&
\begin{bmatrix}
\mathcal{T}(\mathbf{u}_1) & \mathbf{X}\\
\mathbf{X}^H & \mathcal{T}(\mathbf{u}_2)
\end{bmatrix}\succeq0\\
& \mathcal{F}_q(\mathbf{X}) - \mathbf{h}_q\in \mathcal{S}(\mathbf{z}_q), \text{for all}\; q\in\Omega^p,
\end{split}
\end{equation}
where the last constraint enforces consistency with the one-bit data.

Unfortunately, due to the noise, the value of the one-bit sampling may be changed, making the last constraint in (\ref{opt_MinANM}) invalid.
The invalid constraint in (\ref{opt_MinANM}) may result in significant error or even make the problem (\ref{opt_MinANM}) here no feasible solution.
Actually, in the noisy case, the last constraint in (\ref{opt_MinANM}) will become
\begin{equation}
\label{cons_noise}
\mathcal{F}_q(\mathbf{X}) - \mathbf{h}_q + \mathbf{\tilde{w}}_q \in \mathcal{S}(\mathbf{z}_q), \text{for all}\; q\in\Omega^p.
\end{equation}
Since $\mathbf{\tilde{w}}_q$ is random and unknown, it is difficult to directly apply the constraint (\ref{cons_noise}) to our problem.

\vspace{-0.3cm}
\subsection{$\ell_1$-Regularized Atomic-Norm Minimization Method}
In this subsection, we show that the noise $\mathbf{\tilde{w}}_q$ in (\ref{cons_noise}) can be replaced by a sparse perturbation vector.
Then we propose an $\ell_1$-regularized atomic-norm minimization formulation to estimate the low-rank matrix and sparse perturbation simultaneously.

Before proceeding, we give several properties of the expression $\mathbf{a}\in\mathcal{S}(\mathbf{b})$ which is used in the constraint (\ref{cons_noise}).

\begin{lem}\label{lemma1}
For any vectors $\mathbf{a}_1$, $\mathbf{a}_2\in\mathbb{C}^{N}$, if $\mathbf{a}_1\in\mathcal{S}(\mathbf{a}_2)$,
then (1) $\mathbf{a}_2\in\mathcal{S}(\mathbf{a}_1)$ and (2) $\mathbf{a}_1\in\mathcal{S}(\mathbf{a}_1+\mathbf{a}_2)$.
\end{lem}

\begin{lem}\label{lemma2}
For any vectors $\mathbf{a}_1$, $\mathbf{a}_2$, $\mathbf{b}\in\mathbb{C}^{N}$, 
if $\mathbf{a}_1\in\mathcal{S}(\mathbf{b})$ and $\mathbf{a}_2\in\mathcal{S}(\mathbf{b})$,
then $\mathbf{a}_1+\mathbf{a}_2\in\mathcal{S}(\mathbf{b})$.
\end{lem}

\begin{lem}\label{lemma3}
For any vectors $\mathbf{a}_1$, $\mathbf{a}_2$, $\mathbf{b}\in\mathbb{C}^{N}$, 
if $\mathbf{a}_1\in\mathcal{S}(\mathbf{b})$ and $\mathbf{a}_2\in\mathcal{S}(\mathbf{a}_1)$,
then $\mathbf{a}_2\in\mathcal{S}(\mathbf{b})$.
\end{lem}

Lemma~\ref{lemma1}$\sim$Lemma~\ref{lemma3} can be directly proved according to the definition of the expression $\mathbf{a}\in\mathcal{S}(\mathbf{b})$. 
The details of the proof are omitted here for concision.
By applying the above properties, we can derive the following theorem.

\begin{thm}\label{theorem1} 
For any vectors $\mathbf{a}_1$, $\mathbf{a}_2$, $\mathbf{b}\in\mathbb{C}^{N}$, if $\mathbf{a}_1+\mathbf{a}_2\in\mathcal{S}(\mathbf{b})$,
then there exists a vector $\mathbf{a}_3$ satisfying $\mathbf{a}_1+\mathbf{a}_3\in\mathcal{S}(\mathbf{b})$, whose $n$-th element $[\mathbf{a}_3]_n = \Re\{[\mathbf{a}_3]_n\}+j\Im\{[\mathbf{a}_3]_n\}$ is given by
\begin{equation}\label{eqn_Ra3}
\Re\{[\mathbf{a}_3]_n\} = \begin{cases}
0     &\Re\{[\mathbf{a}_2]_n\}\Re\{[\mathbf{a}_1]_n\}\geq0 \\
0     &|\Re\{[\mathbf{a}_1]_n\}|\geq|\Re\{[\mathbf{a}_2]_n|\},\\
\Re\{[\mathbf{a}_2]_n\}  & \text{else}.\\
\end{cases}
\end{equation}

\begin{equation}\label{eqn_Ia3}
\Im\{[\mathbf{a}_3]_n\} = \begin{cases}
0     &\Im\{[\mathbf{a}_2]_n\}\Im\{[\mathbf{a}_1]_n\}\geq0 \\
0     &|\Im\{[\mathbf{a}_1]_n\}|\geq|\Im\{[\mathbf{a}_2]_n|\},\\
\Im\{[\mathbf{a}_2]_n\}  & \text{else}.\\
\end{cases}
\end{equation}
\end{thm}

\begin{IEEEproof}
If $\Re\{[\mathbf{a}_2]_n\}\Re\{[\mathbf{a}_1]_n\}\geq0$, we get $\Re\{[\mathbf{a}_1]_n\}\in\mathcal{S}(\Re\{[\mathbf{a}_2]_n\})$.
Then, by applying Lemma~\ref{lemma1}, we have $\Re\{[\mathbf{a}_1]_n\}\in\mathcal{S}(\Re\{[\mathbf{a}_1]_n\}+\Re\{[\mathbf{a}_2]_n\})$.
According to Lemma~\ref{lemma3}, we can derive that $\Re\{[\mathbf{a}_1]_n\}\in\mathcal{S}(\Re\{[\mathbf{b}]_n\})$.

If $|\Re\{[\mathbf{a}_1]_n\}|\geq|\Re\{[\mathbf{a}_2]_n\}|$, we have $\Re\{[\mathbf{a}_1]_n\}(\Re\{[\mathbf{a}_1]_n\}+\Re\{[\mathbf{a}_2]_n\})\geq0$, i.e., $\Re\{[\mathbf{a}_1]_n\}\in\mathcal{S}(\Re\{[\mathbf{a}_1]_n\}+\Re\{[\mathbf{a}_2]_n\})$.
Thus, we can also derive that $\Re\{[\mathbf{a}_1]_n\}\in\mathcal{S}(\Re\{[\mathbf{b}]_n\})$.

For the imaginary part, we can derive similar results.
\end{IEEEproof}

Theorem~\ref{theorem1} demonstrates that, for any $\mathbf{a}_1+\mathbf{a}_2\in\mathcal{S}(\mathbf{b})$, 
there exists a vector $\mathbf{a}_3$ with $\|\mathbf{a}_3\|_0\leq\|\mathbf{a}_2\|_0$ such that $\mathbf{a}_1+\mathbf{a}_3\in\mathcal{S}(\mathbf{b})$.
For the constraint (\ref{cons_noise}), we can find a vector $\mathbf{p}_q$ according to Theorem~\ref{theorem1} such that
\begin{equation}
\label{cons_perturb}
\mathcal{F}_q(\mathbf{X}) - \mathbf{h}_q + \mathbf{p}_q \in \mathcal{S}(\mathbf{z}_q), \text{for all}\; q\in\Omega^p,
\end{equation}
where $\|\mathbf{p}_q\|_0\leq\|\mathbf{\tilde{w}_q}\|_0$. 
To distinguish it from the noise $\mathbf{\tilde{w}_q}$, we call $\mathbf{p}_q$ the perturbation vector here.
Since the noise $\mathbf{\tilde{w}_q}$ is random, it might be possible that only a small portion of the elements of $\mathbf{p}_q$ is nonzero. 
In Figure 1, we plot the average percentage of nonzero elements in $\mathbf{p}_q$ with respect to different signal-to-noise ratios (SNRs) when Gaussian noise exists in the signal.
It is observed that, when $\text{SNR}\geq 0$ dB, the percentage of nonzero elements in $\mathbf{p}_q$ is no more than $15\%$, i.e., $\mathbf{p}_q$ is sparse. 
Therefore, it is possible to recover the low-rank matrix $\mathbf{X}$ and the sparse perturbation $\mathbf{p}_q$ simultaneously.

\begin{figure}[!t]
\centering
\includegraphics[width=2in]{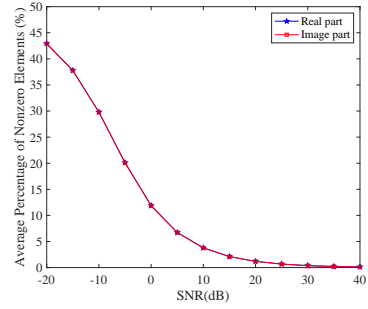}
\caption{Average percentage of nonzero elements in $\mathbf{p}_q$ with respect to different SNRs.}
\label{SignChangeRate}
\end{figure}

To enforce the sparsity of the perturbation vector, we formulate the following one-bit $\ell_1$-regularized atomic-norm minimization (1b-ANM- L1) problem:
\begin{equation}
\label{eqn_RecX1}
\begin{split}
\min_{\mathbf{u}_1,\mathbf{u}_2,\mathbf{X},\mathbf{p}_q} & \Tr(\mathcal{T}(\mathbf{u}_1))+\Tr(\mathcal{T}(\mathbf{u}_2))+\lambda\sum_{q\in\Omega^p}\|\mathbf{p}_q\|_{1},\\
\text{s.t.}\quad&
\begin{bmatrix}
\mathcal{T}(\mathbf{u}_1) & \mathbf{X}\\
\mathbf{X}^H & \mathcal{T}(\mathbf{u}_2)
\end{bmatrix}\succeq0\\
& \mathcal{F}_q(\mathbf{X}) - \mathbf{h}_q + \mathbf{p}_q \in \mathcal{S}(\mathbf{z}_q), \text{for all}\; q\in\Omega^p,
\end{split}
\end{equation}
where $\lambda>0$ is a regularization parameter.
This problem is convex and can be directly solved by using the convex optimization toolbox CVX \cite{cvx}. 

After solving the problem (\ref{eqn_RecX1}), there are several methods to get the angle and Doppler frequency parameters.
One method is to apply the conventional two-dimensional frequency estimation method \cite{Kay-TASSP1990, Hua-TSP1992} once the low-rank matrix $\mathbf{X}$ is recovered.
Another method is to explore the two Toeplitz matrices $\mathcal{T}(\mathbf{u}_1)$ and $\mathcal{T}(\mathbf{u}_2)$ recovered in (\ref{eqn_RecX1}).
By performing Vandermonde decomposition of the two Toeplitz matrices, we can get two sets of $K$ frequencies, which correspond to the angles and Doppler frequencies, respectively. 
Then a simple pairing method, as shown in \cite{Xi:ICASSP-2017,Tian-ICASSP2017}, can be implemented to recover the $K$ angle-Doppler pairs.
After we get the estimates of the angle-Doppler pairs, the reflection coefficients of the $K$ targets can also be estimated.
If the number of targets is unknown, we can apply the Bayesian information criterion (BIC) to select the model order.
 The recently developed one-bit Bayesian information criterion (1bBIC) [39] can be used to determine the model order directly from one-bit measurements.
According to Appendix A, we know that $K = \rank(\mathbf{H})$.
Thus, we can also exploit this property to determine the model order.

\section{An ADMM-based Iterative Algorithm}
\label{Sec:ADMM}
Although the convex optimization toolbox can solve the problem (\ref{eqn_RecX1}), it does not scale well when solving large-scale problems.
To accelerate the computation, in this section we develop an iterative algorithm to solve the 1b-ANM-L1 problem via the alternating direction method of multipliers (ADMM) \cite{Boyd:ADMM-11}.

The key to applying the ADMM method is to write out the augmented Lagrangian function of the 1b-ANM-L1 problem.
However, due to the last constraint in (\ref{eqn_RecX1}), it is not direct to write out the augmented Lagrangian function.
In this section, we introduce a new auxiliary vector $\mathbf{b}_q = |\Re\{\mathcal{F}_q(\mathbf{X})-\mathbf{h}_q+\mathbf{p}_q\}|+j|\Im\{\mathcal{F}_q(\mathbf{X})-\mathbf{h}_q+\mathbf{p}_q\}|$, where $|\cdot|$ denotes the element-wise absolute value.
Then we can derive the following expression:
\begin{equation}
\label{eqn_zq}
\mathbf{z}_q\tilde{\odot} \mathbf{b}_q = \mathcal{F}_q(\mathbf{X})-\mathbf{h}_q+\mathbf{p}_q,
\end{equation}
where $\tilde{\odot}$ denotes the complex-valued element-wise product of vectors or matrices, i.e., 
$\mathbf{a}\tilde{\odot}\mathbf{b}=\Re\{\mathbf{a}\}\odot\Re\{\mathbf{b}\}+j\Im\{\mathbf{a}\}\odot\Im\{\mathbf{b}\}$.

First, by applying (\ref{eqn_zq}), we rewrite the problem (\ref{eqn_RecX1}) as
\begin{equation}
\label{eqn_ADMM1}
\begin{split}
\min_{\mathcal{X},\mathcal{B},\mathcal{P}}\quad
& MN[\mathbf{u_1}]_1+Q[\mathbf{u_2}]_1+\lambda\sum_{q\in\Omega^p}\|\mathbf{p}_q\|_{1}\\
& +\frac \mu 2\sum_{q\in\Omega^p}\|\mathcal{F}_q(\mathbf{X})-\mathbf{h}_q+\mathbf{p}_q-\mathbf{z}_q\tilde{\odot}\mathbf{b}_q\|_2^2\\
& +\mathbf{I}_\infty(\mathbf{H}\succeq0),\\
\text{ s.t.}\qquad &\mathbf{H}=\begin{bmatrix}
\mathcal{T}(\mathbf{u}_1) & \mathbf{X}\\
\mathbf{X}^H & \mathcal{T}(\mathbf{u}_2)
\end{bmatrix},\\
\end{split}
\end{equation}
where $\mathcal{X}=\{\mathbf{u}_1,\mathbf{u}_2,\mathbf{X}\}$, $\mathcal{P}=\{\mathbf{p}_q\}_{q\in\Omega^p}$, 
    and $\mathcal{B}=\{\mathbf{b}_q\}_{q\in\Omega^p}$ denote the unknown parameters to be optimized,
$\mu$ is a regularization parameter, 
and $\mathbf{I}_\infty(\cdot)$ denotes an indicator function that is 0 if the condition in the bracket is true, and infinity otherwise.

Then the augmented Lagrangian function of the problem (\ref{eqn_ADMM1}) can be expressed as
\begin{equation}
  \label{eqn_Lag1}
  \begin{split}
&\mathcal{L}_{\rho}(\mathcal{X},\mathcal{P},\mathcal{B},\boldsymbol{\Lambda},\mathbf{H})\\
=&MN[\mathbf{u_1}]_1+Q[\mathbf{u_2}]_1+\lambda\sum_{q\in\Omega^p}\|\mathbf{p}_q\|_{1}\\
+&\frac \mu 2\sum_{q\in\Omega^p}\|\mathcal{F}_q(\mathbf{X})-\mathbf{h}_q+\mathbf{p}_q-\mathbf{z}_q\tilde{\odot}\mathbf{b}_q\|_2^2\\
+&\mathbf{I}_\infty(\mathbf{H}\succeq0)+\frac {\rho}2\left\|\mathbf{H}-\begin{bmatrix}
\mathcal{T}(\mathbf{u}_1) & \mathbf{X}\\
\mathbf{X}^H & \mathcal{T}(\mathbf{u}_2)
\end{bmatrix}\right\|_F^2\\
+&\left\langle\boldsymbol{\Lambda},\mathbf{H}-\begin{bmatrix}
\mathcal{T}(\mathbf{u}_1) & \mathbf{X}\\
\mathbf{X}^H & \mathcal{T}(\mathbf{u}_2)
\end{bmatrix}\right\rangle,\\
\end{split}
\end{equation}
where $\boldsymbol{\Lambda}$ is the Lagrangian multiplier, and $\rho>0$ is the penalty parameter. 

According to \cite{Boyd:ADMM-11}, at the $(l+1)$-th iteration, the ADMM update takes the following steps:
\begin{equation}
  \label{eqn_ADMMupdate1}
\{\mathcal{X}^{l+1},\mathcal{P}^{l+1}\}=
\arg\min_{\mathcal{X},\mathcal{P}}\mathcal{L}_{\rho}(\mathcal{X},\mathcal{P},\mathcal{B}^l,\boldsymbol{\Lambda}^l,\mathbf{H}^l),
\end{equation}
\begin{equation}
\label{eqn_update_bq}
\begin{split}
	\mathbf{b}_q^{l+1} =& \left|\Re\{\mathcal{F}_q(\mathbf{X}^{l+1})-\mathbf{h}_q+\mathbf{p}_q^{l+1}\}\right|\\
	+&j\left|\Im\{\mathcal{F}_q(\mathbf{X}^{l+1})-\mathbf{h}_q+\mathbf{p}_q^{l+1}\}\right|,
\end{split}
\end{equation}
\begin{equation}
  \label{eqn_ADMMupdate2}
\mathbf{H}^{l+1}=
\arg\min_{\mathbf{H}\geq0}\mathcal{L}_{\rho}(\mathcal{X}^{l+1},\mathcal{P}^{l+1},\mathcal{B}^{l+1},\boldsymbol{\Lambda}^l,\mathbf{H}),
\end{equation}
\begin{equation}
  \label{eqn_ADMMupdate3}
\boldsymbol{\Lambda}^{l+1}=
\boldsymbol{\Lambda}^{l}+\rho\left(\mathbf{H}^{l+1}-\begin{bmatrix}
\mathcal{T}(\mathbf{u}_1^{l+1}) & \mathbf{X}^{l+1}\\
{(\mathbf{X}^{l+1}})^H & \mathcal{T}(\mathbf{u}_2^{l+1})
\end{bmatrix}\right).
\end{equation}

Since the updates in (\ref{eqn_update_bq}) and (\ref{eqn_ADMMupdate3}) is explicit, 
we derive the updates of (\ref{eqn_ADMMupdate1}) and (\ref{eqn_ADMMupdate2}) in details.
It is noted that the update in (\ref{eqn_update_bq}) is based on the definition of $\mathbf{b}_q$.

\vspace{-0.3cm}
\subsection{Update of $\mathcal{X}^{l}$, $\mathcal{P}^{l}$, and $\mathbf{H}^{l}$}

Since it is difficult to solve the problem (\ref{eqn_ADMMupdate1}) directly, we update $\mathcal{X}$ and $\mathcal{P}$ separately by solving two sub-problems:
\begin{equation}
  \label{eqn_updateX}
\mathcal{X}^{l+1}=
\arg\min_{\mathcal{X}}\mathcal{L}_{\rho}(\mathcal{X},\mathcal{P}^l,\mathcal{B}^l,\boldsymbol{\Lambda}^l,\mathbf{H}^l),
\end{equation}
\begin{equation}
  \label{eqn_updateP}
\mathcal{P}^{l+1}=
\arg\min_{\mathcal{P}}\mathcal{L}_{\rho}(\mathcal{X}^l,\mathcal{P},\mathcal{B}^l,\boldsymbol{\Lambda}^l,\mathbf{H}^l).
\end{equation}
For convenience, we introduce the following partitions of the matrices $\mathbf{H}^{l}$ and $\boldsymbol{\Lambda}^{l}$:
\begin{equation}
  \label{eqn_QPartition}
\mathbf{H}^{l}=\begin{bmatrix}
\mathbf{H}_1^{l} & \mathbf{H}_{\mathbf{X}}^{l}\\
(\mathbf{H}_{\mathbf{X}}^{l})^H & \mathbf{H}_2^{l}
\end{bmatrix},
\end{equation}
\begin{equation}
  \label{eqn_LambdaPartition}
\boldsymbol{\Lambda}^{l}=\begin{bmatrix}
\boldsymbol{\Lambda}_1^{l} & \boldsymbol{\Lambda}_{\mathbf{X}}^{l}\\
(\boldsymbol{\Lambda}_{\mathbf{X}}^{l})^H & \boldsymbol{\Lambda}_2^{l}
\end{bmatrix},
\end{equation}
where $\mathbf{H}_1^{l}$ and $\boldsymbol{\Lambda}_1^{l}$ are $MN\times MN$ matrices, 
$\mathbf{H}_{\mathbf{X}}^{l}$ and $\boldsymbol{\Lambda}_{\mathbf{X}}^{l}$ are $MN\times Q$ matrices, 
and $\mathbf{H}_2^{l}$ and $\boldsymbol{\Lambda}_2^{l}$ are $Q\times Q$ matrices, respectively.

Then we compute the derivatives of $\mathcal{L}_{\rho}(\mathcal{X},\mathcal{P},\mathcal{B},\boldsymbol{\Lambda}^l,\mathbf{H}^l)$ with respect to $\mathbf{X}$ and the elements of $\mathbf{u}_1$ and $\mathbf{u}_2$.
The results are given by (\ref{eqn_DerivX})$\sim$(\ref{eqn_Derivu2}) at the top of this page,
where $\mathbf{m}_{q,n}^H$ is the $n$-th row of $\mathbf{M}_q$, 
$\mathbf{z}_q^l=\mathbf{p}_q^l-\mathbf{h}_q-\mathbf{z}_q\tilde{\odot}\mathbf{b}_q^l$,
and $\textrm{Tr}_n(\cdot)$ outputs the trace of the $n$-th sub-diagonal of the input matrix.

\newcounter{mytempeqncnt}
\begin{figure*}[!t]
\normalsize
\begin{equation}
  \label{eqn_DerivX}
  \begin{split}
\triangledown_{\mathbf{X}}\mathcal{L}_{\rho}&=\mu\sum_{q\in\Omega^p}\sum_{n=1}^{LR}\mathbf{m}_{q,n}{\mathbf{e}_q^H}\left (\langle\mathbf{X},{\mathbf{m}_{q,n}{\mathbf{e}_q^H}}\rangle+[\mathbf{z}_q^l]_n\right)-2\boldsymbol{\Lambda}_{\mathbf{X}}^l+2\rho\left(\mathbf{X}-\mathbf{H}_{\mathbf{X}}^l\right),
\end{split}
\end{equation}

\begin{equation}
  \label{eqn_Derivu1}
   \triangledown_{[\mathbf{u}_1]_n}\mathcal{L}_{\rho}=\begin{cases}
   MN+{\rho}MN[\mathbf{u}_1]_1-\textrm{Tr}\left(\rho\mathbf{H}_1^l+\boldsymbol{\Lambda}_1^l\right), &n=1;\\
   {\rho}(MN-n+1)[\mathbf{u}_1]_n-\textrm{Tr}_n\left(\rho\mathbf{H}_1^l+\boldsymbol{\Lambda}_1^l\right),&n=2,\cdots,MN,
   \end{cases}
\end{equation}

\begin{equation}
  \label{eqn_Derivu2}
   \triangledown_{[\mathbf{u}_2]_n}\mathcal{L}_{\rho}=\begin{cases}
   Q+{\rho}Q[\mathbf{u}_2]_1-\textrm{Tr}\left(\rho\mathbf{H}_2^l+\boldsymbol{\Lambda}_2^l\right), &n=1;\\
   {\rho}(Q-n+1)[\mathbf{u}_2]_n-\textrm{Tr}_n\left(\rho\mathbf{H}_2^l+\boldsymbol{\Lambda}_2^l\right),&n=2,\cdots,Q.
   \end{cases}
\end{equation}

\hrulefill
\vspace*{4pt}
\end{figure*}

By setting the derivatives to be 0, $\mathbf{X}^{l+1}$, $\mathbf{u}_1^{l+1}$ and $\mathbf{u}_2^{l+1}$ can be updated by
\begin{equation}
  \label{eqn_UpdateX}
   \mathbf{x}^{l+1}_q=\begin{cases}
\left(\mu\mathbf{M}_{q}^H\mathbf{M}_{q}+2\rho\mathbf{I}_{MN}\right)^{-1}\\
\times\left((2\boldsymbol{\Lambda}_{\mathbf{X}}^l+2\rho\mathbf{H}_{\mathbf{X}}^l)\mathbf{e}_q-\mu\mathbf{M}_{q}^H\mathbf{z}_q^l\right), &q\in\Omega^p,\\
(\rho^{-1}\boldsymbol{\Lambda}_{\mathbf{X}}^l+\mathbf{H}_{\mathbf{X}}^l)\mathbf{e}_q, &q\notin\Omega^p,
   \end{cases}
\end{equation}
\begin{equation}
  \label{eqn_UpdateU1}
   [\mathbf{u}_1]^{l+1}_n=\begin{cases}
\frac 1{\rho MN}\textrm{Tr}\left(\rho\mathbf{H}_1^l+\boldsymbol{\Lambda}_1^l\right)-\frac{1}{\rho}, &n=1,\\
\frac 1{\rho(MN-n+1)}\textrm{Tr}_n\left(\rho\mathbf{H}_1^l+\boldsymbol{\Lambda}_1^l\right),&n=2,\cdots,MN,
   \end{cases}
\end{equation}
\begin{equation}
  \label{eqn_UpdateU2}
   [\mathbf{u}_2]^{l+1}_n=\begin{cases}
\frac 1{\rho Q}\textrm{Tr}\left(\rho\mathbf{H}_2^l+\boldsymbol{\Lambda}_2^l\right)-\frac{1}{\rho}, &n=1,\\
\frac 1{\rho(Q-n+1)}\textrm{Tr}_n\left(\rho\mathbf{H}_2^l+\boldsymbol{\Lambda}_2^l\right),&n=2,\cdots,Q,
   \end{cases}
\end{equation}
where $\mathbf{x}^{l+1}_q$ denotes the $q$-th column of the matrix $\mathbf{X}^{l+1}$.

According to (\ref{eqn_Lag1}), the update of $\mathbf{p}_q$, $q\in\Omega^p$, is equivalent to solving the following problem:
\begin{equation}
\label{eqn_update_p}
\mathbf{p}_q^{l+1} = \arg\min_{\mathbf{p}_q}\frac 1 2\|\mathcal{F}_q(\mathbf{X}^l)-\mathbf{h}_q+\mathbf{p}_q-\mathbf{z}_q\tilde{\odot}\mathbf{b}_q^l\|_2^2 + \frac\lambda\mu\|\mathbf{p}_q\|_{1}.
\end{equation}
Therefore, $\mathbf{p}_q$ can be updated by
\begin{equation}
\label{eqn_update_pq}
\mathbf{p}_q^{l+1}=\text{Prox}_{\frac\lambda\mu}\left(\mathbf{z}_q\tilde{\odot}\mathbf{b}_q^l+\mathbf{h}_q-\mathcal{F}_q(\mathbf{X}^l)\right),
\end{equation}
where $\text{Prox}_{\lambda}(\mathbf{x})$ is the proximal operator \cite{Maleki-TIT2013}, in which each element of $\text{Prox}_{\lambda}(\mathbf{x})$, denoted as $[\text{Prox}_{\lambda}(\mathbf{x})]_n$, is given by
\begin{equation}
\label{eqn_prox}
[\text{Prox}_{\lambda}(\mathbf{x})]_n=
\begin{cases}
[\mathbf{x}]_n-\lambda\frac{[\mathbf{x}]_n}{|[\mathbf{x}]_n|}, &|[\mathbf{x}]_n|>\lambda,\\
0, &|[\mathbf{x}]_n|\leq\lambda.
\end{cases}
\end{equation}

Finally, the update of $\mathbf{H}^l$ is equivalent to solving the following problem:
\begin{equation}
  \label{eqn_UpdateH}
\mathbf{H}^{l+1}=
\arg\min_{\mathbf{H}\succeq0}\left\|\mathbf{H}-\begin{bmatrix}
\mathcal{T}(\mathbf{u}_1^{l+1}) & \mathbf{X}^{l+1}\\
{(\mathbf{X}^{l+1}})^H & \mathcal{T}(\mathbf{u}_2^{l+1})
\end{bmatrix}+\frac1\rho{\boldsymbol{\Lambda}^l}\right\|_F^2.
\end{equation}
The solution is to project the matrix $\begin{bmatrix}
\mathcal{T}(\mathbf{u}_1^{l+1}) & \mathbf{X}^{l+1}\\
{(\mathbf{X}^{l+1}})^H & \mathcal{T}(\mathbf{u}_2^{l+1})
\end{bmatrix}-\frac1\rho\boldsymbol{\Lambda}^l$ onto the positive definite cone.
This projection can be accomplished by setting all the negative eigenvalues of the matrix to zero.

\vspace{-0.3cm}
\subsection{Summary of the ADMM-based Algorithm}
The proposed ADMM-based iterative algorithm is summarized as Algorithm 1. 
It is noted that $\mathbf{b}_q$ is initialized to be $\mathbf{1}$ rather than $\mathbf{0}$.
If $\mathbf{b}_q$ were initialized to be $\mathbf{0}$, the term $\mathbf{z}_q\tilde{\odot} \mathbf{b}_q$ used in (\ref{eqn_UpdateX}) would be zero,
and the information provided by $\mathbf{z}_q$ could not be used in the update process.

\begin{algorithm}[htb]
\caption{The ADMM-based 1b-ANM-L1 Algorithm}
\begin{algorithmic}[1]
\REQUIRE $\mathbf{u}_1^0=\mathbf{0}$, $\mathbf{u}_2^0=\mathbf{0}$, $\mathbf{p}_q^0=\mathbf{0}$, $\mathbf{b}_q^0=\mathbf{1}$,
$\mathbf{X}^0=\mathbf{0},\mathbf{H}^0=\mathbf{0},\boldsymbol{\Lambda}^0=\mathbf{0}$;
\ENSURE While the stop condition is not satisfied, do;

\STATE Update $\mathbf{X}^l$ using (\ref{eqn_UpdateX});
\STATE Update $\mathbf{u}_1^l$ and $\mathbf{u}_2^l$ using (\ref{eqn_UpdateU1}) and (\ref{eqn_UpdateU2}), respectively;
\STATE Update $\mathbf{p}_q^l$ and $\mathbf{b}_q^l$ using (\ref{eqn_update_pq}) and (\ref{eqn_update_bq}), respectively;
\STATE Update $\mathbf{H}^l$ by solving (\ref{eqn_UpdateH});
\STATE Update $\boldsymbol{\Lambda}^l$ using (\ref{eqn_ADMMupdate3}).

\end{algorithmic}
\end{algorithm}

\vspace{-0.25cm}
\section{Performance Analysis based on Cram$\acute{\text{e}}$r-Rao Bound}
The Cram$\acute{\text{e}}$r-Rao bound (CRB) provides a theoretical limit on the variance of any unbiased parameter estimator.
In this section, we study the effect of one-bit sampling on parameter estimation by analyzing its CRB. 
Furthermore, based on the derived CRB, two different threshold strategies for one-bit sampling are also discussed.

In our problem, the set of deterministic but unknown parameters to be estimated is $\boldsymbol{\theta}=\{\vartheta_k,\upsilon_k,r_k,\phi_k\}_{k=1}^K$,
where $r_k$ and $\phi_k$ denote the magnitude and phase of the reflection coefficient $\beta_k=r_k e^{j\phi}$, respectively.
We assume that the number of targets, the noise level, and the transmitted signals are known.
Since the transmitted signals are known, the operator $\mathcal{F}_q(\cdot)$ is deterministic.
Each entry of $\mathbf{\tilde{w}}_q$ is assumed to be i.i.d. complex Gaussian, i.e., 
$[\mathbf{\tilde{w}}_q]_n \sim \mathcal{CN}(0,\sigma^2)$.

\vspace{-0.25cm}
\subsection{CRB for Unquantized Data}\label{Subsec:CRB_Unq}
We first establish the CRB for unquantized data.
According to (\ref{eqn_ysubq2}), the probability density function (PDF) of the unquantized data $\mathbf{\tilde{y}}_q$, denoted as $p(\mathbf{\tilde{y}}_q|\boldsymbol{\theta})$, is expressed as
\begin{equation}
\label{eqn_PDFyq}
p(\mathbf{\tilde{y}}_q|\boldsymbol{\theta})=\frac{1}{(\pi\sigma^2)^{LR}}\exp\left(-\frac{\|\mathbf{\tilde{y}}_q-\mathcal{F}_q(\mathbf{X})\|_2^2}{\sigma^2}\right).
\end{equation}
According to the definition of the Fisher Information Matrix (FIM) of $\mathbf{\tilde{y}}_q$, denoted by $\mathbf{\tilde{I}}_q(\boldsymbol{\theta})\in\mathbb{R}^{4K\times 4K}$, we have
\begin{equation}
\label{eqn_Iyq}
\begin{split}
\mathbf{\tilde{I}}_q(\boldsymbol{\theta}) &= \mathbb{E}\left[\left(\frac{\partial\log p(\mathbf{\tilde{y}}_q|\boldsymbol{\theta})}{\partial\boldsymbol{\theta}}\right)
\left(\frac{\partial\log p(\mathbf{\tilde{y}}_q|\boldsymbol{\theta})}{\partial\boldsymbol{\theta}}\right)^T\right]\\
&=\frac{2}{\sigma^2}\sum_{n=1}^{LR}\Re\left\{\left(\frac{\partial[\mathcal{F}_q(\mathbf{X})]^\ast_n}{\partial\boldsymbol{\theta}}\right)\left(\frac{\partial[\mathcal{F}_q(\mathbf{X})]_n}{\partial\boldsymbol{\theta}}\right)^T\right\}.
\end{split}
\end{equation}
Let $\mathbf{r}_q$ and $\mathbf{i}_q$ be the real and imaginary parts of $\mathcal{F}_q(\mathbf{X})$, i.e., $\mathbf{r}_q=\Re\{\mathcal{F}_q(\mathbf{X})\}$ and $\mathbf{i}_q=\Im\{\mathcal{F}_q(\mathbf{X})\}$.
Then (\ref{eqn_Iyq}) can be equivalently expressed as
\begin{equation}
\label{eqn_FIMyq2}
\mathbf{\tilde{I}}_q(\boldsymbol{\theta})=\sum_{n=1}^{LR}\left[\mathbf{\tilde{I}}_{q,n}^{R}(\boldsymbol{\theta})+
\mathbf{\tilde{I}}_{q,n}^{I}(\boldsymbol{\theta})\right],
\end{equation}
where
\begin{equation}
\label{eqn_IqnR_Unq}
\mathbf{\tilde{I}}_{q,n}^{R}(\boldsymbol{\theta})=\frac{2}{\sigma^2}\left(\frac{\partial[\mathbf{r}_q]_n}{\partial\boldsymbol{\theta}}\right)\left(\frac{\partial[\mathbf{r}_q]_n}{\partial\boldsymbol{\theta}}\right)^T,
\end{equation}
\begin{equation}
\label{eqn_IqnI_Unq}
\mathbf{\tilde{I}}_{q,n}^{I}(\boldsymbol{\theta})=\frac{2}{\sigma^2}\left(\frac{\partial[\mathbf{i}_q]_n}{\partial\boldsymbol{\theta}}\right)\left(\frac{\partial[\mathbf{i}_q]_n}{\partial\boldsymbol{\theta}}\right)^T.
\end{equation}

Considering the $P$ pulses sent out by the transmitter, the FIM of the set of received unquantized data $\{\mathbf{\tilde{y}}_q\}_{q\in\Omega^p}$ is
\begin{equation}
\mathbf{\tilde{I}}(\boldsymbol{\theta})=\sum_{q\in\Omega^p}\mathbf{\tilde{I}}_q(\boldsymbol{\theta}).
\end{equation}
Then the CRB on the variance of the unbiased estimate of the $i$-th parameter is the $i$-th diagonal element of the inverse $(\mathbf{\tilde{I}}(\boldsymbol{\theta}))^{-1}$.
Generally, the FIM depends on the values of the parameters. 

\vspace{-0.3cm}
\subsection{CRB for 1-bit Quantized Data}
For the one-bit sampled data given in (\ref{eqn_Qyq}), 
the probability mass function (PMF) of $\mathbf{z}_q$, denoted as $p(\mathbf{z}_q|\boldsymbol{\theta})$, 
is expressed as
\begin{equation}
\label{eqn_PMF}
p(\mathbf{z}_q|\boldsymbol{\theta})=\prod_{n=1}^{LR}p(\Re\{[\mathbf{z}_q]_n\}|\boldsymbol{\theta})
p(\Im\{[\mathbf{z}_q]_n\}|\boldsymbol{\theta}),
\end{equation}
where 
\begin{equation}
\begin{split}
p(\Re\{[\mathbf{z}_q]_n\}|\boldsymbol{\theta}) &= \mathbb{P}(\Re\{[\mathbf{z}_q]_n\}=1|\boldsymbol{\theta})^{\frac{1+\Re\{[\mathbf{z}_q]_n\}}{2}}\\
&\times \mathbb{P}(\Re\{[\mathbf{z}_q]_n\}=-1|\boldsymbol{\theta})^{\frac{1-\Re\{[\mathbf{z}_q]_n\}}{2}},
\end{split}
\end{equation}
and $p(\Im\{[\mathbf{z}_q]_n\}|\boldsymbol{\theta})$ is expressed similarly by replacing $\Re\{[\mathbf{z}_q]_n\}$ with $\Im\{[\mathbf{z}_q]_n\}$.

Let $\mathbf{h}^r_q$ and $\mathbf{h}^i_q$ be the real and imaginary parts of $\mathbf{h}_q$.
Since we have $\Re\{[\mathbf{\tilde{y}}_q-\mathbf{h}_q]_n\} \sim \mathcal{N}([\mathbf{r}_q-\mathbf{h}^r_q]_n,\frac{1}{2}\sigma^2)$,
we can derive that
\begin{equation}
\label{eqn_Probzq1}
\begin{split}
\mathbb{P}(\Re\{[\mathbf{z}_q]_n\}=1|\boldsymbol{\theta})&=\mathbb{P}(\Re\{[\mathbf{\tilde{y}}_q-\mathbf{h}_q]_n\}\geq0|\boldsymbol{\theta})\\
&=\Phi\left(\frac{[\mathbf{r}_q-\mathbf{h}^r_q]_n}{\sigma}\right),
\end{split}
\end{equation}
\begin{equation}
\label{eqn_Probzq}
\begin{split}
\mathbb{P}(\Re\{[\mathbf{z}_q]_n\}=-1|\boldsymbol{\theta})&=\mathbb{P}(\Re\{[\mathbf{\tilde{y}}_q-\mathbf{h}_q]_n\}<0|\boldsymbol{\theta})\\
&=1-\Phi\left(\frac{[\mathbf{r}_q-\mathbf{h}^r_q]_n}{\sigma}\right),
\end{split}
\end{equation}
where $\Phi(x)=\frac{1}{\sqrt{\pi}}\int_{-\infty}^{x}e^{-t^2}dt$.
Similar results can be derived for $\Im\{[\mathbf{z}_q]_n\}$.

By applying the results given in \cite{Fu2018-TSP}, the FIM of the 1-bit quantized data $\mathbf{z}_q$, denoted as $\mathbf{I}_q(\boldsymbol{\theta})$, can be stated as
\begin{equation}
\mathbf{I}_q(\boldsymbol{\theta})=\sum_{n=1}^{LR}\left[\mathbf{I}_{q,n}^{R}(\boldsymbol{\theta})+
\mathbf{I}_{q,n}^{I}(\boldsymbol{\theta})\right],
\end{equation}
where
\begin{equation}
\label{eqn_IqnR_1b}
\mathbf{I}_{q,n}^{R}(\boldsymbol{\theta})=\frac{2}{\sigma^2}\omega\left(\frac{[\mathbf{r}_q-\mathbf{h}^r_q]_n}{\sigma}\right)
\left(\frac{\partial[\mathbf{r}_q]_n}{\boldsymbol{\theta}}\right)
\left(\frac{\partial[\mathbf{r}_q]_n}{\boldsymbol{\theta}}\right)^T,
\end{equation}
\begin{equation}
\label{eqn_IqnI_1b}
\mathbf{I}_{q,n}^{I}(\boldsymbol{\theta})=\frac{2}{\sigma^2}\omega\left(\frac{[\mathbf{i}_q-\mathbf{h}^i_q]_n}{\sigma}\right)
\left(\frac{\partial[\mathbf{i}_q]_n}{\boldsymbol{\theta}}\right)
\left(\frac{\partial[\mathbf{i}_q]_n}{\boldsymbol{\theta}}\right)^T,
\end{equation}
with $\omega(x)=\frac{\exp(-2x^2)}{2\pi\Phi(x)[1-\Phi(x)]}$.

With $P$ pulses, the FIM becomes
\begin{equation}
\mathbf{I}(\boldsymbol{\theta}) = \sum_{q\in\Omega^p}\mathbf{I}_q(\boldsymbol{\theta}).
\end{equation}
Then the CRB can be determined by the diagonal elements of the inverse $(\mathbf{I}(\boldsymbol{\theta}))^{-1}$.

Comparing the results in (\ref{eqn_IqnR_1b}) and (\ref{eqn_IqnI_1b}) with those in (\ref{eqn_IqnR_Unq}) and (\ref{eqn_IqnI_Unq}), 
it is seen that the FIM of one-bit sampled data is a weighted version of the FIM of unquantized data, 
with the weights of $\omega\left(\frac{[\mathbf{r}_q-\mathbf{h}^r_q]_n}{\sigma}\right)$ for the real part 
and $\omega\left(\frac{[\mathbf{i}_q-\mathbf{h}^i_q]_n}{\sigma}\right)$ for the imaginary part.
Therefore, the weight function $\omega(x)$, which is plotted in Fig. \ref{weightfunction}, has an important effect on the FIM.
Considering Fig. \ref{weightfunction}, we make the following comments on the FIM $\mathbf{I}(\boldsymbol{\theta})$ of one-bit sampled data:
\begin{enumerate}[(i)]
\item The upper bound of $\mathbf{I}(\boldsymbol{\theta})$ is $\frac{2}{\pi}\mathbf{\tilde{I}}(\boldsymbol{\theta})\approx 0.64\mathbf{\tilde{I}}(\boldsymbol{\theta})$, i.e., $\mathbf{I}(\boldsymbol{\theta})\preceq\frac{2}{\pi}\mathbf{\tilde{I}}(\boldsymbol{\theta})$.
However, in finite SNR scenarios, it is required that $\mathbf{r}_q=\mathbf{h}^r_q$ and $\mathbf{i}_q=\mathbf{h}^i_q$ to achieve the upper bound, which is impossible in practical applications.
The upper bound also proves that one-bit sampling incurs at least a 2 dB information loss \cite{Papadopoulos-TIT2001}.

\item Given a signal and a threshold, the FIM $\mathbf{I}(\boldsymbol{\theta})$ is closer to its upper bound when the noise variance $\sigma^2$ is larger, i.e., when the SNR is lower. 
When the SNR increases, the gap between the FIM $\mathbf{I}(\boldsymbol{\theta})$ and its upper bound increases.
It is also proved that the parameter estimation performance with one-bit sampling is closer to that with high-bit sampling in the low SNR regime.
However, in the high SNR regime, a performance gap separates one-bit sampling and high-bit sampling.

\item When no threshold is applied, the value of the weight function $\omega(x)$ is inversely proportional to the SNR, i.e., the larger the SNR, the lower the weight function.
This relationship leads to significant information loss in the high SNR regime.
Thus we prefer to apply the non-zero threshold to one-bit sampling, especially in the high SNR regime.
Clearly, designing an appropriate threshold strategy is an important problem in one-bit sampling.
\end{enumerate}
\vspace{-0.3cm}
\subsection{Threshold Strategies}
In this subsection, two different threshold strategies are considered, and the effects of the threshold strategies on the FIM $\mathbf{I}(\boldsymbol{\theta})$ are discussed.

\subsubsection{Random Uniform Threshold}
The first threshold strategy is to let each element of the threshold $[\mathbf{h}^r_q]_n$ and $[\mathbf{h}^i_q]_n$ uniformly distributed between $h_{max}$ and $h_{min}$, i.e., $[\mathbf{h}^r_q]_n,[\mathbf{h}^i_q]_n\sim\mathcal{U}[h_{min},h_{max}]$.

\begin{figure}[!t]
\centering
\includegraphics[width=2in]{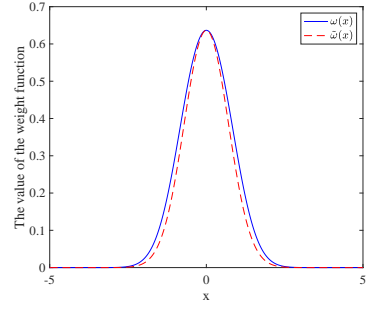}
\caption{The weight function $\omega(x)$.}
\label{weightfunction}
\end{figure}

To simplify our analysis, we approximate the weight function $\omega(x)$ as $\tilde{\omega}(x)=\frac{2}{\pi}\exp(-x^2)$.
The curve of $\tilde{\omega}(x)$, shown in Fig. \ref{weightfunction}, approximates the weight function well.
Then the expectation of the FIM $\mathbf{I}_{q,n}^{R}(\boldsymbol{\theta})$ with respect to the threshold value $[\mathbf{h}^r_q]_n$ can be computed as
\begin{equation}\label{eqn_EI}
\begin{split}&\mathbb{E}\{\mathbf{I}_{q,n}^{R}(\boldsymbol{\theta})\}\\
=&\frac{2}{\sigma^2}\mathbb{E}\left\{\omega\left(\frac{[\mathbf{r}_q-\mathbf{h}^r_q]_n}{\sigma}\right)\right\}
\left(\frac{\partial[\mathbf{r}_q]_n}{\boldsymbol{\theta}}\right)
\left(\frac{\partial[\mathbf{r}_q]_n}{\boldsymbol{\theta}}\right)^T\\
\approx&\frac{2}{\sigma^2}\mathbb{E}\left\{\tilde{\omega}\left(\frac{[\mathbf{r}_q-\mathbf{h}^r_q]_n}{\sigma}\right)\right\}
\left(\frac{\partial[\mathbf{r}_q]_n}{\boldsymbol{\theta}}\right)
\left(\frac{\partial[\mathbf{r}_q]_n}{\boldsymbol{\theta}}\right)^T\\
=&\frac{2}{\sigma^2}\frac{2\sigma}{\sqrt{\pi}\Delta_h}\left(\tilde{\Phi}_{min}-\tilde{\Phi}_{max}\right)
\left(\frac{\partial[\mathbf{r}_q]_n}{\boldsymbol{\theta}}\right)
\left(\frac{\partial[\mathbf{r}_q]_n}{\boldsymbol{\theta}}\right)^T,
\end{split}
\end{equation}
where $\Delta_h=h_{max}-h_{min}$, and the last line comes from
\begin{equation}
\begin{split}
\mathbb{E}\left\{\tilde{\omega}\left(\frac{[\mathbf{r}_q-\mathbf{h}^r_q]_n}{\sigma}\right)\right\} &= \frac{1}{\Delta_h}\int_{h_{min}}^{h_{max}}\tilde{\omega}\left(\frac{[\mathbf{r}_q]_n-h}{\sigma}\right) dh\\
&=\frac{2\sigma}{\sqrt{\pi}\Delta_h}\left(\tilde{\Phi}_{min}-\tilde{\Phi}_{max}\right),
\end{split}
\end{equation}
with $\tilde{\Phi}_{min}=\Phi\left(\frac{[\mathbf{r}_q]_n-h_{min}}{\sigma}\right)$ and $\tilde{\Phi}_{max}=\Phi\left(\frac{[\mathbf{r}_q]_n-h_{max}}{\sigma}\right)$.

The expectation of the FIM $\mathbf{I}_{q,n}^{I}(\boldsymbol{\theta})$ can also be computed similarly.
We omit the results here for concision.
According to the mean value theorem, there exists a $\xi\in[\frac{[\mathbf{r}_q]_n-h_{max}}{\sigma},\frac{[\mathbf{r}_q]_n-h_{min}}{\sigma}]$ such that
\begin{equation}
\frac{\left(\tilde{\Phi}_{min}-\tilde{\Phi}_{max}\right)}{\Delta_h}=\frac{\Phi'(\xi)}{\sigma}=\frac{e^{-\xi^2}}{\sqrt{\pi}\sigma}.
\end{equation}
If $0\in[\frac{[\mathbf{r}_q]_n-h_{max}}{\sigma},\frac{[\mathbf{r}_q]_n-h_{min}}{\sigma}]$, $\mathbb{E}\{\mathbf{I}_{q,n}^{R}(\boldsymbol{\theta})\}$ is upper bounded by $\frac{2}{\pi}\tilde{\mathbf{I}}(\boldsymbol{\theta})$, i.e., $\mathbb{E}\{\mathbf{I}(\boldsymbol{\theta})\}\preceq\frac{2}{\pi}\tilde{\mathbf{I}}(\boldsymbol{\theta})$.
In practice, since we do not know the exact value of $[\mathbf{r}_q]_n$, we can set $h_{min}$ and $h_{max}$ as the minimum and maximum values of $[\mathbf{r}_q]_n$, respectively, such that $0\in[\frac{[\mathbf{r}_q]_n-h_{max}}{\sigma},\frac{[\mathbf{r}_q]_n-h_{min}}{\sigma}]$ for every $n$.

\subsubsection{Random Gaussian Threshold}
The second threshold strategy is to let each element of the threshold $[\mathbf{h}^r_q]_n$ be an i.i.d. Gaussian random variable with mean $[\mathbf{r}_q]_n$ and variance $\sigma_r^2$,
i.e., $[\mathbf{h}^r_q]_n\sim\mathcal{N}([\mathbf{r}_q]_n,\sigma_r^2)$.
Similarly, $[\mathbf{h}^i_q]_n\sim\mathcal{N}([\mathbf{i}_q]_n,\sigma_i^2)$.

Then we can derive that
\begin{equation}
\label{Eqn_Eomega}
\begin{split}
&\mathbb{E}\left\{\tilde{\omega}\left(\frac{[\mathbf{r}_q-\mathbf{h}^r_q]_n}{\sigma}\right)\right\}\\
=&\frac{2}{\pi}\int_{-\infty}^{\infty} \exp(-x^2)\times\frac{\sigma}{\sqrt{2\pi}\sigma_r}\exp(-\frac{x^2\sigma^2}{2\sigma_r^2})dx\\
=&\frac{2\sigma}{\pi\sqrt{2\sigma_r^2+\sigma^2}}.
\end{split}
\end{equation}
The expectation of the FIM $\mathbf{I}_{q,n}^{R}(\boldsymbol{\theta})$ with respect to the threshold value $[\mathbf{h}^r_q]_n$ can be approximated as
\begin{equation}
\begin{split}
&\mathbb{E}\{\mathbf{I}_{q,n}^{R}(\boldsymbol{\theta})\}\\
\approx&\frac{2}{\sigma^2}\mathbb{E}\left\{\tilde{\omega}\left(\frac{[\mathbf{r}_q-\mathbf{h}^r_q]_n}{\sigma}\right)\right\}
\left(\frac{\partial[\mathbf{r}_q]_n}{\boldsymbol{\theta}}\right)
\left(\frac{\partial[\mathbf{r}_q]_n}{\boldsymbol{\theta}}\right)^T\\
=&\frac{2}{\sigma^2}\frac{2\sigma}{\pi\sqrt{2\sigma_r^2+\sigma^2}}
\left(\frac{\partial[\mathbf{r}_q]_n}{\boldsymbol{\theta}}\right)
\left(\frac{\partial[\mathbf{r}_q]_n}{\boldsymbol{\theta}}\right)^T.
\end{split}
\end{equation}
A similar result can be derived for the FIM $\mathbf{I}_{q,n}^{I}(\boldsymbol{\theta})$.
It is noted that the expectation of the weight function depends on only the noise variance $\sigma$ and the threshold variance $\sigma_r$ or $\sigma_i$.
If $\sigma_r^2 = \kappa\sigma^2$, then (\ref{Eqn_Eomega}) becomes $\frac{2}{\pi\sqrt{2\kappa+1}}$,
which is independent of the noise variance $\sigma$.
This property is useful in improving the performance of one-bit sampling in the high SNR regime.
In practice, \textit{a priori} estimates of $\mathbf{r}_q$ and $\mathbf{i}_q$ can be set as the mean values of the random Gaussian threshold.

The thresholds we considered here are assumed to be infinite precision, 
i.e., they can take any value. 
Actually, due to the finite bit-depth of digital-to-analog converters (DACs), 
the thresholds have to be quantized to finite precision, i.e., they can take only a set of discrete values.

\vspace{-0.3cm}
\subsection{CRB for the Case with Unknown Noise Variance}
In some practical applications, the noise level in the received signal is unknown.
Thus, it is necessary to analyze the CRB under the assumption of unknown noise variance.
In this case, the unknown parameters to be estimated become $\boldsymbol{\theta}$ and $\sigma$.

For unquantized data, we can apply the similar analysis in \cite{Stoica:CRB-TASSP-1989} and derive that the FIM can be expressed as
\begin{equation}\label{Eqn_FIM_unq_noise}
	\begin{bmatrix}
		\frac{4LRP}{\sigma^2} &0\\
		0 &\tilde{\mathbf{I}}(\boldsymbol{\theta})\\
	\end{bmatrix}.
\end{equation}
It means that the CRB on the variance of the unbiased estimate of $\sigma$ is $\frac{\sigma^2}{4LRP}$, and the CRB on the variance of the unbiased estimate of $\boldsymbol{\theta}$ is same as that shown in Section \ref{Subsec:CRB_Unq}.
The details of the above derivation are omitted here.

For 1-bit quantized data, the FIM becomes
\begin{equation}\label{Eqn_FIM_unq_noise_1b}
	\begin{bmatrix}
		\mathbf{I}(\sigma) &\mathbf{I}(\sigma,\boldsymbol{\theta})^T\\
		\mathbf{I}(\sigma,\boldsymbol{\theta}) &\mathbf{I}(\boldsymbol{\theta})
	\end{bmatrix},
\end{equation}
where $\mathbf{I}(\sigma,\boldsymbol{\theta})=\left[\mathbf{I}(\sigma,\theta_1),\cdots,\mathbf{I}(\sigma,\theta_{4K})\right]^T$ for $\theta_1,\cdots, \theta_{4K}\in\boldsymbol{\theta}$ with
\begin{equation}
\begin{split}
	\mathbf{I}(\sigma,\theta_i)=&-\frac{2}{\sigma^2}\sum_{q\in\Omega^p}\sum_{n=1}^{LR}\omega\left(\frac{[\mathbf{r}_q-\mathbf{h}^r_q]_n}{\sigma}\right)\frac{[\mathbf{r}_q-\mathbf{h}^r_q]_n}{\sigma}\frac{\partial[\mathbf{r}_q]_n}{\partial\theta_i}\\
	-&\frac{2}{\sigma^2}\sum_{q\in\Omega^p}\sum_{n=1}^{LR}\omega\left(\frac{[\mathbf{i}_q-\mathbf{h}^i_q]_n}{\sigma}\right)\frac{[\mathbf{i}_q-\mathbf{h}^i_q]_n}{\sigma}\frac{\partial[\mathbf{i}_q]_n}{\partial\theta_i},
\end{split}
\end{equation}
and
\begin{equation}
\begin{split}
	\mathbf{I}(\sigma)=&\frac{2}{\sigma^2}\sum_{q\in\Omega^p}\sum_{n=1}^{LR}\omega\left(\frac{[\mathbf{r}_q-\mathbf{h}^r_q]_n}{\sigma}\right)\frac{[\mathbf{r}_q-\mathbf{h}^r_q]_n^2}{\sigma^2}\\
	+&\frac{2}{\sigma^2}\sum_{q\in\Omega^p}\sum_{n=1}^{LR}\omega\left(\frac{[\mathbf{i}_q-\mathbf{h}^i_q]_n}{\sigma}\right)\frac{[\mathbf{i}_q-\mathbf{h}^i_q]_n^2}{\sigma^2}.
\end{split}
\end{equation}
Therefore, the CRBs on the variance of the unbiased estimate of $\sigma$ and $\boldsymbol{\theta}$ are the diagonal elements of the inverse of the above Fisher information matrix.
The details of the above derivation are given in Appendix B.

\begin{figure*}[!t]
\centering
\subfigure[]{
\begin{minipage}[t]{0.32\linewidth}
\centering
\includegraphics[width=1.8in]{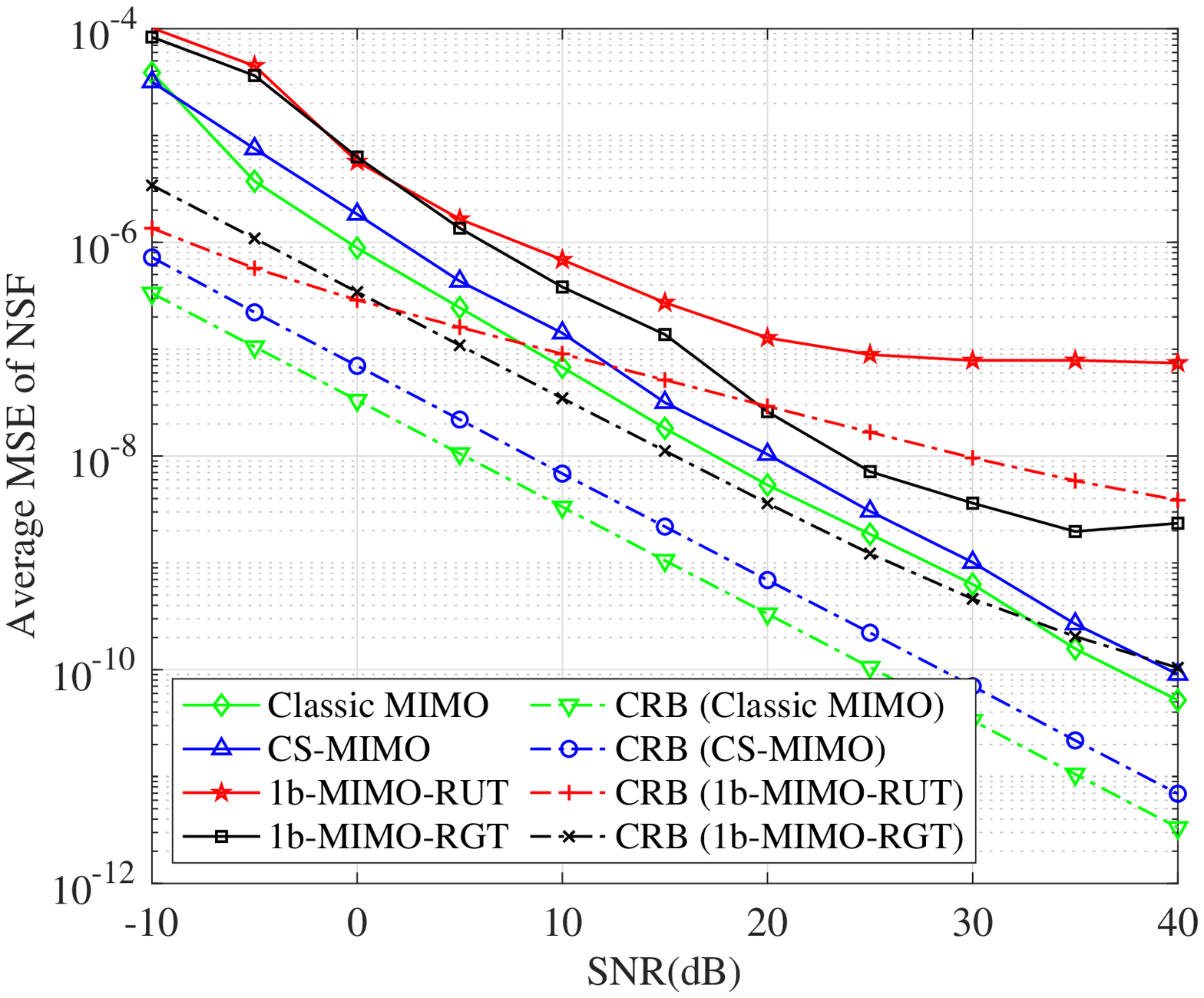}
\end{minipage}%
}%
\subfigure[]{
\begin{minipage}[t]{0.32\linewidth}
\centering
\includegraphics[width=1.8in]{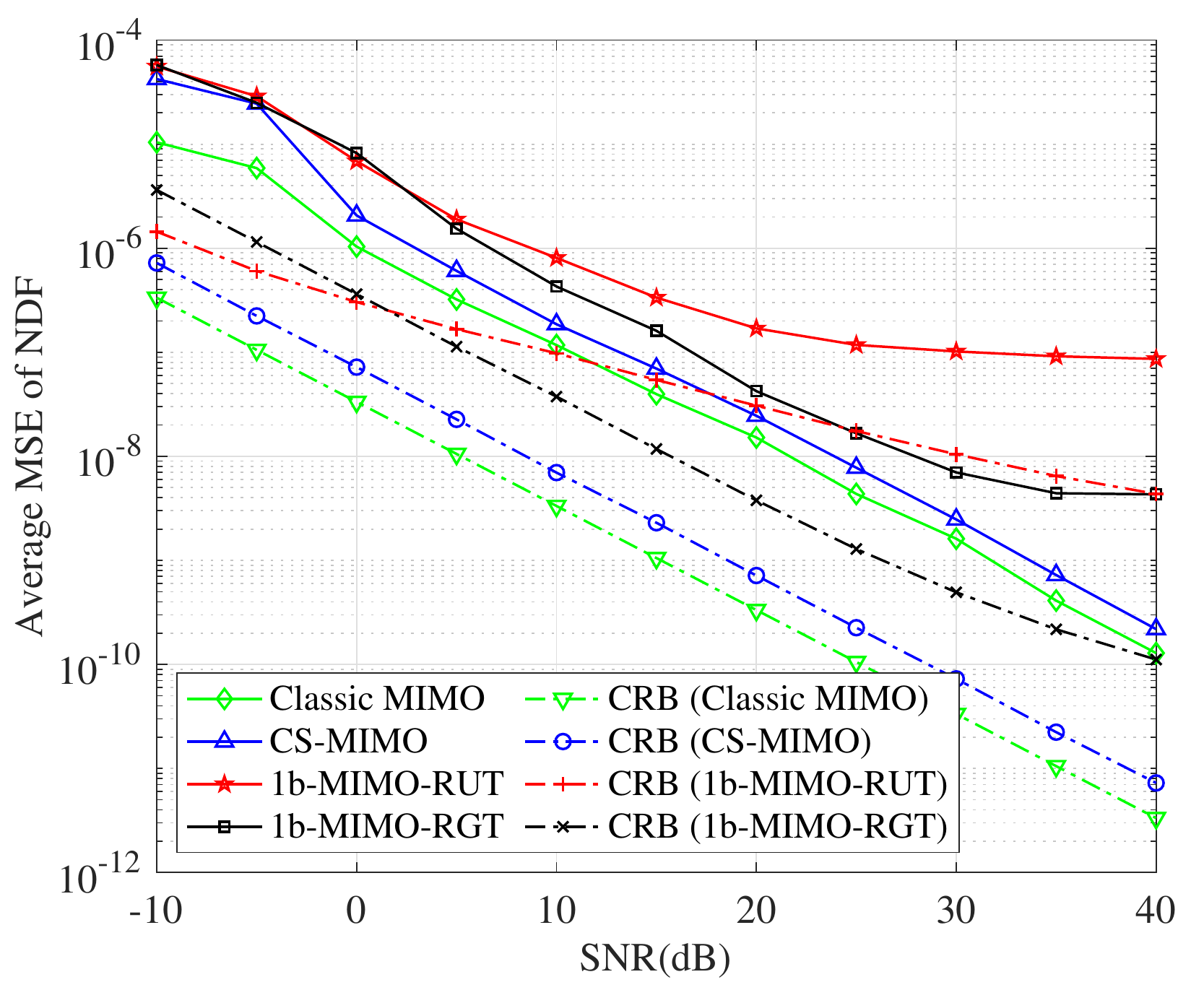}
\end{minipage}%
}%
\subfigure[]{
\begin{minipage}[t]{0.32\linewidth}
\centering
\includegraphics[width=1.8in]{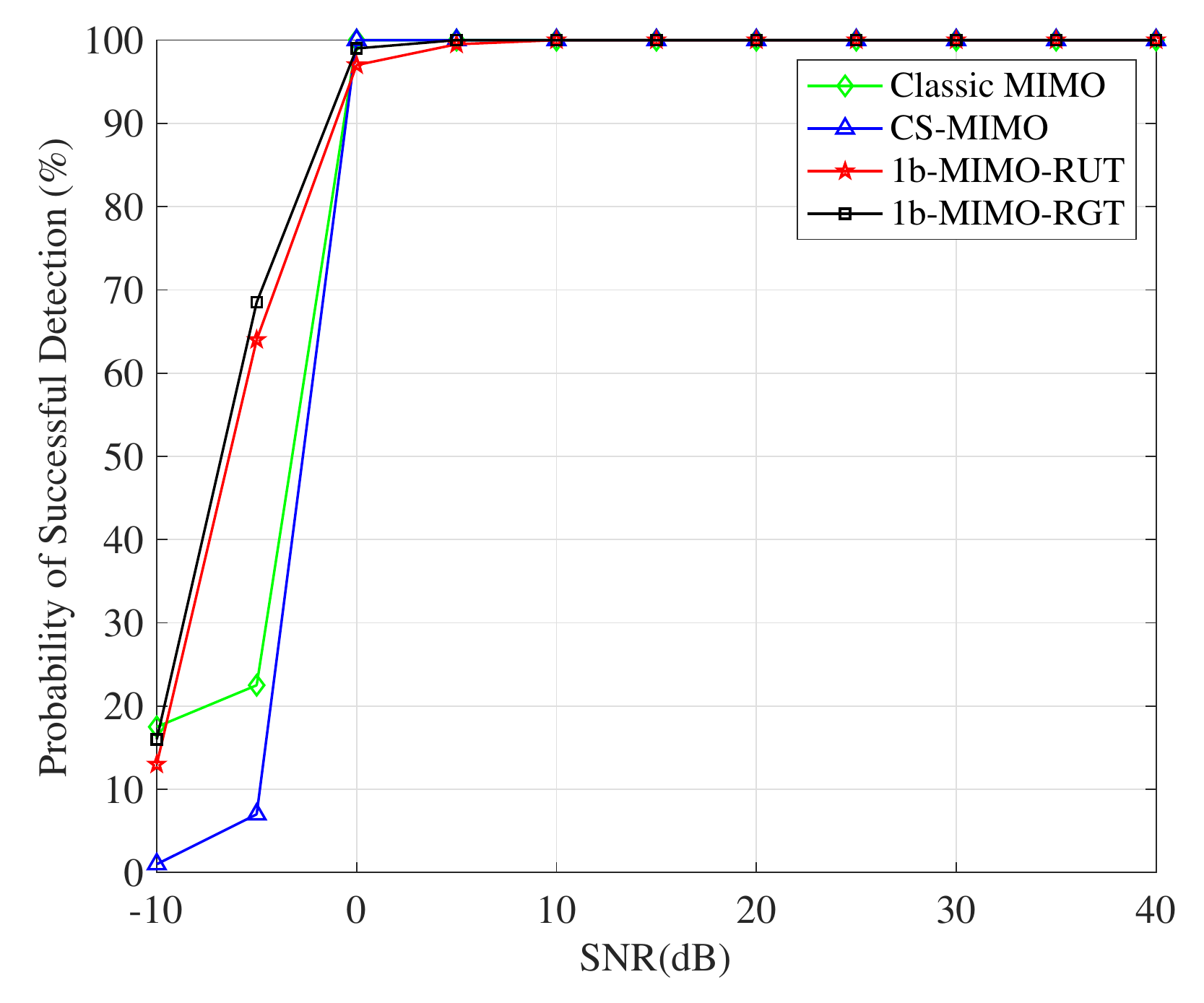}
\end{minipage}
}\\%
\subfigure[]{
\begin{minipage}[t]{0.32\linewidth}
\centering
\includegraphics[width=1.8in]{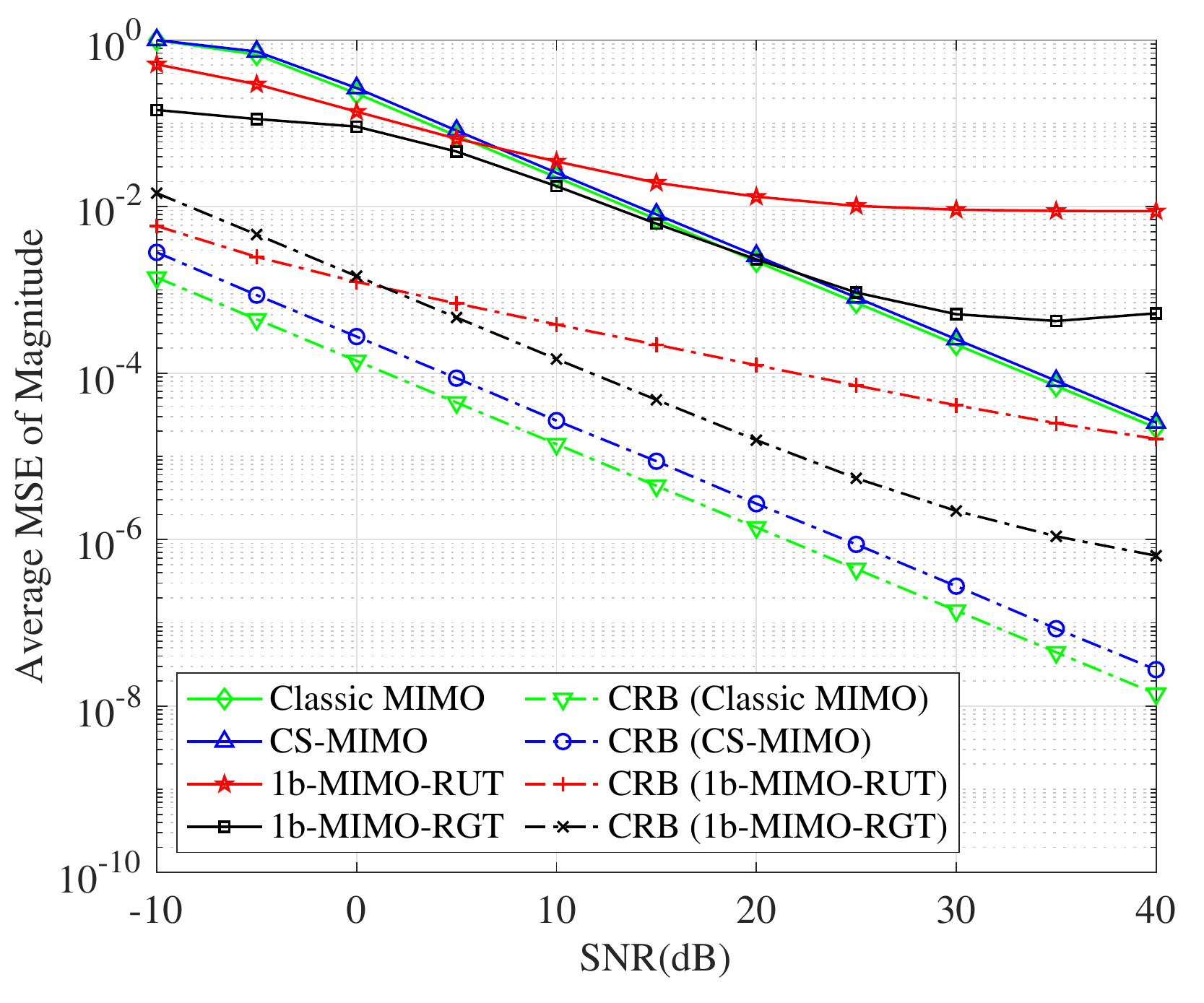}
\end{minipage}%
}%
\subfigure[]{
\begin{minipage}[t]{0.32\linewidth}
\centering
\includegraphics[width=1.8in]{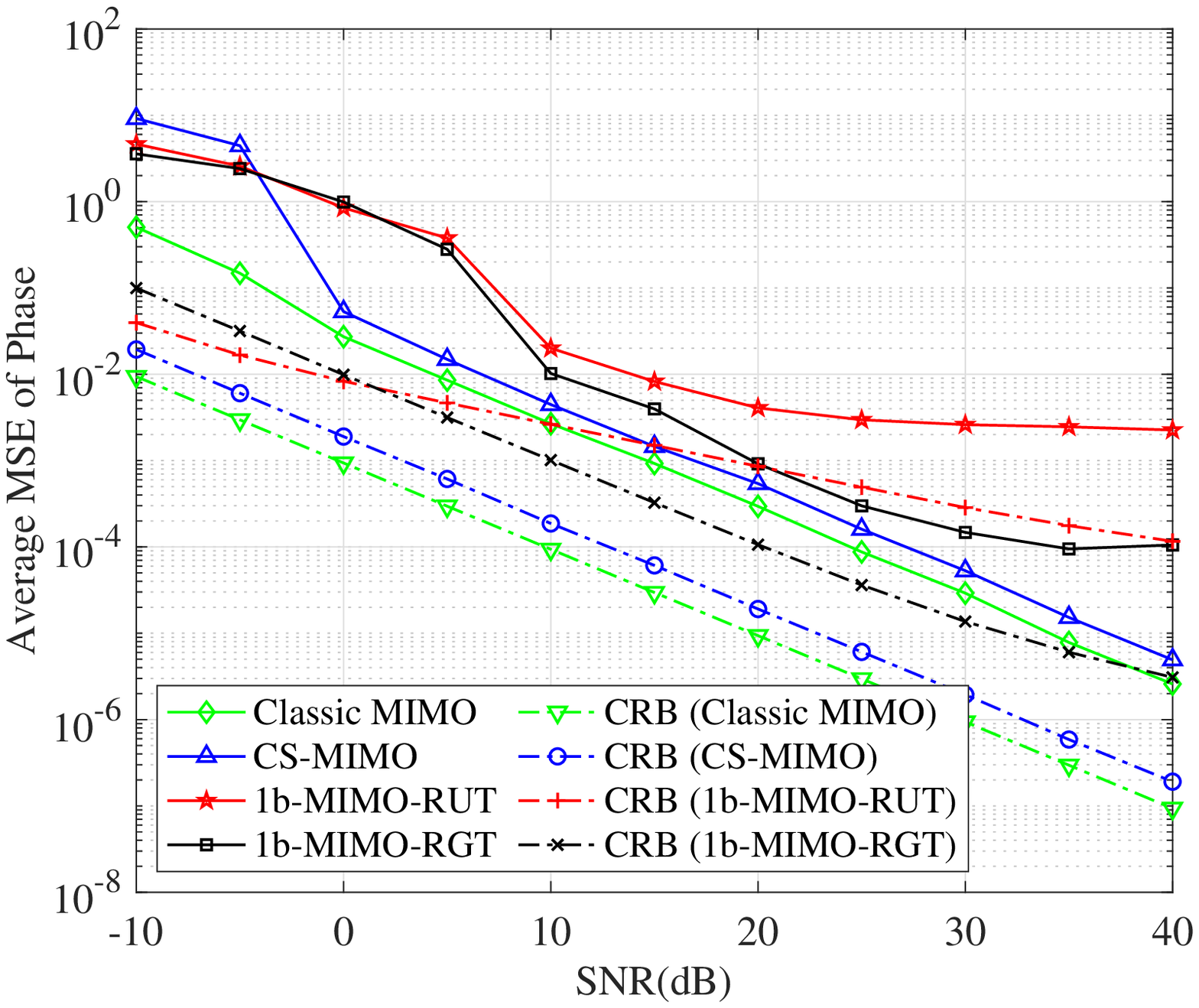}
\end{minipage}%
}%
\subfigure[]{
\begin{minipage}[t]{0.32\linewidth}
\centering
\includegraphics[width=1.8in]{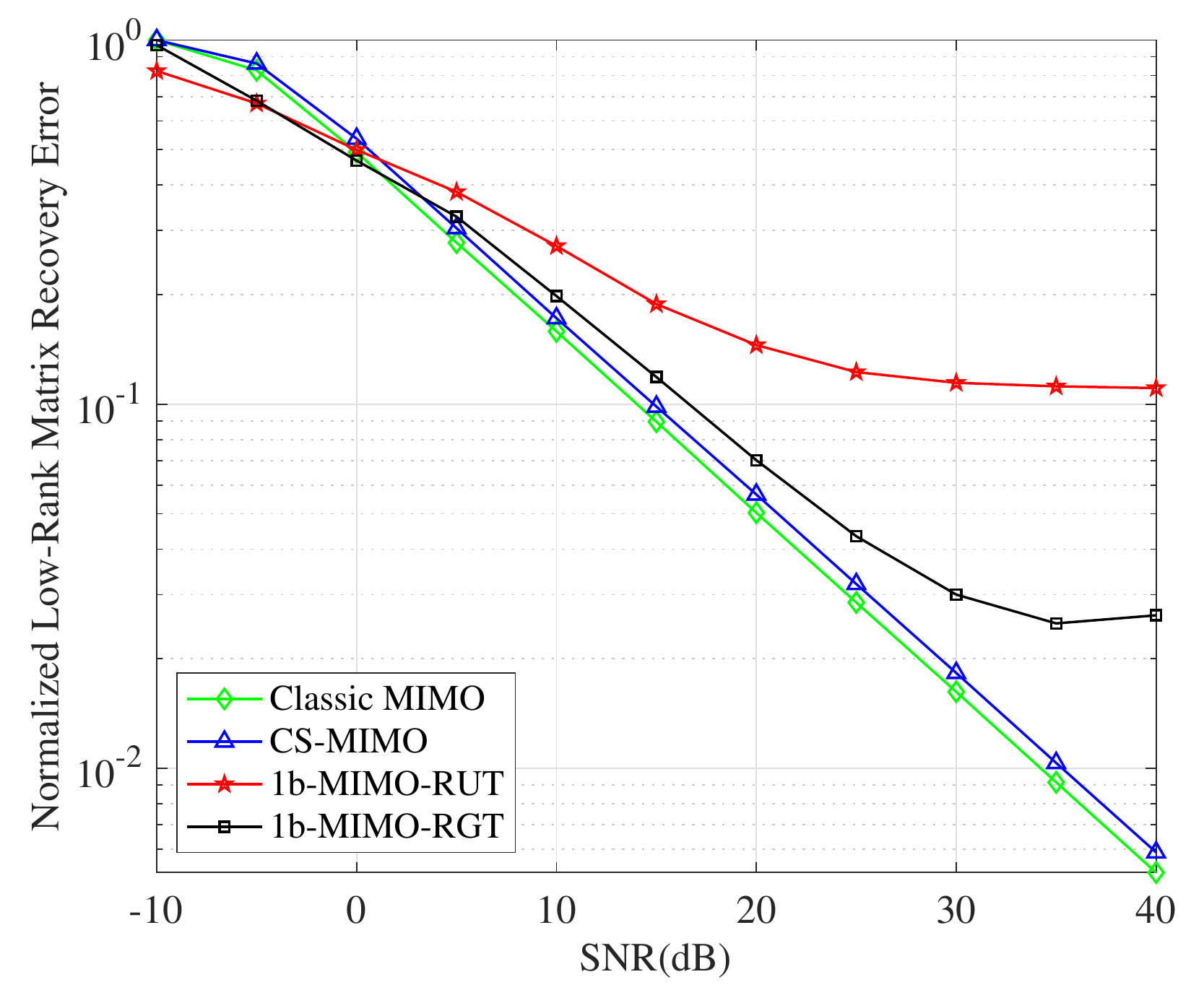}
\end{minipage}%
}%
\centering
\caption{Estimation Performance with respect to SNR when $K=4$. (a) Average MSE of NSF vs. SNR; (b) Average MSE of NDF vs. SNR; (c) Probability of successful detection vs. SNR; (d) Average MSE of magnitude vs. SNR; (e) Average MSE of phase vs. SNR; (f) Normalized low-rank matrix recovery error vs. SNR.}
\end{figure*}

\section{Numerical Experiments}
In this section, we present numerical experiments to illustrate our proposed 1b-MIMO radar, as well as the 1b-ANM-L1 algorithm.
As a comparison, two kinds of high-bit quantized MIMO radars are considered.
One is a CS-MIMO radar that performs CS in the spatial and Doppler domains but applies 16-bit sampling in the temporal domain.
The other is the classic MIMO radar with 16-bit sampling and without any CS techniques.
\vspace{-0.3cm}
\subsection{Simulation Setup}

Throughout the simulations, we consider a MIMO radar with $N=6$ receive antennas and $M=6$ transmit antennas. 
One CPI includes $Q=36$ pulses.
A set of quasi-orthogonal waveforms with code length $L=64$ is used as the transmitted waveforms if not specified.
For the 1b-MIMO radar and CS-MIMO radar, $T=4$ transmit antennas and $R=5$ receive antennas are randomly chosen. 
During one CPI, $P=20$ pulses are randomly chosen to send out pulses.
With these settings, the total amount of data generated by the 1b-MIMO radar is only $3\%$ as great as the data from the classic MIMO radar, and only $6\%$ of the total data from the CS-MIMO radar.
In the 1b-MIMO radar, both the RUT and RGT strategies are implemented to generate the threshold.
In the RUT strategy, the threshold is randomly distributed between the minimum value and maximum value of the received signal.
In the RGT strategy, the mean of the threshold is the estimates of $\mathbf{r}_q$ and $\mathbf{i}_q$ provided by the 1b-MIMO radar using RUT $\footnote{Here is to show the best achievable performance of the RGT strategy. A more practical way is to use a portion of the data to get \textit{a priori} estimates.}$, and the variance is $\sigma_r^2=\sigma_i^2=5\sigma^2$. 
All these thresholds are quantized to 12-bits before applying them to 1-bit sampling.
The 1b-MIMO radar and the CS-MIMO radar solve problem (\ref{eqn_RecX1}) and problem (\ref{opt_MinANMunq}), respectively.
The angle and Doppler frequency are estimated from the recovered low-rank matrix.
For the classic MIMO radar, the MEMP method \cite{Hua-TSP1992} is applied to perform the two-dimensional parameter estimation.
To reduce the noise, atomic norm denoising \cite{Bhaskar-TSP2013} is used after matched filtering. 
To evaluate the estimation performance, the average mean-squared errors (MSEs) are computed from 200 Monte Carlo runs.
In each trial, the noise and threshold are realized independently.

For the proposed ADMM-based 1b-ANM-L1 algorithm, the regularization parameters $\mu$, $\lambda$, and $\rho$ are set as $2/(1+e^{-0.25*\text{SNR}})$, $50$, and $0.5$, respectively.
When the normalized error between two iterations is less than $1e-6$ or the number of iterations reaches $10^3$, the ADMM-based iteration is stopped.

\vspace{-0.3cm}
\subsection{Simulation Results}
We first study the estimation accuracy of the 1b-MIMO radar for various SNR values.
In this experiment, $K=4$ targets are considered, with normalized spatial frequency (NSF) and normalized Doppler frequency (NDF) pairs of $\{-0.1594,0.3805\}$,$\{-0.4480,0.1274\}$,$\{0.3036,-0.2268\}$, and $\{0.3036,-0.4330\}$.
The magnitudes of the reflection coefficients are fixed to be 1, and the phases of the reflection coefficients are randomly distributed between $[0,2\pi]$.
Note that the last two targets have the same NSFs, i.e., they are located in the same direction.
The average MSEs of the NSF and NDF, as well as the probability of successful detection (PSD) with respect to SNR, are shown in Fig. 3(a)$\sim$(c), respectively.
A successful detection is declared if the estimation errors of the NSF and NDF are both less than $1/MN=1/Q=1/36$, i.e., one resolution bin.
Only the successful detection trials are taken into account in the computation of the average MSEs.
As shown, in the low SNR regime ($\text{SNR}\leq 10\text{dB}$), the performance of the 1b-MIMO radar is close to the CS-MIMO radar and classic MIMO radar. 
And in the high SNR regime, the 1b-MIMO radar can achieve MSEs of the NSF and NDF as low as $10^{-9}\sim10^{-7}$.
However, the performance gap between the 1b-MIMO radar and those high-bit quantized rivals increases as the SNR increases.
This trend is in accordance with our analysis in Section VI.
As shown in Fig. 3(c), the 1b-MIMO radar can achieve a higher PSD in the low SNR regime than its high-bit rivals.

\begin{figure}[!t]
\label{ccc}
\centering
\includegraphics[width=1.8in]{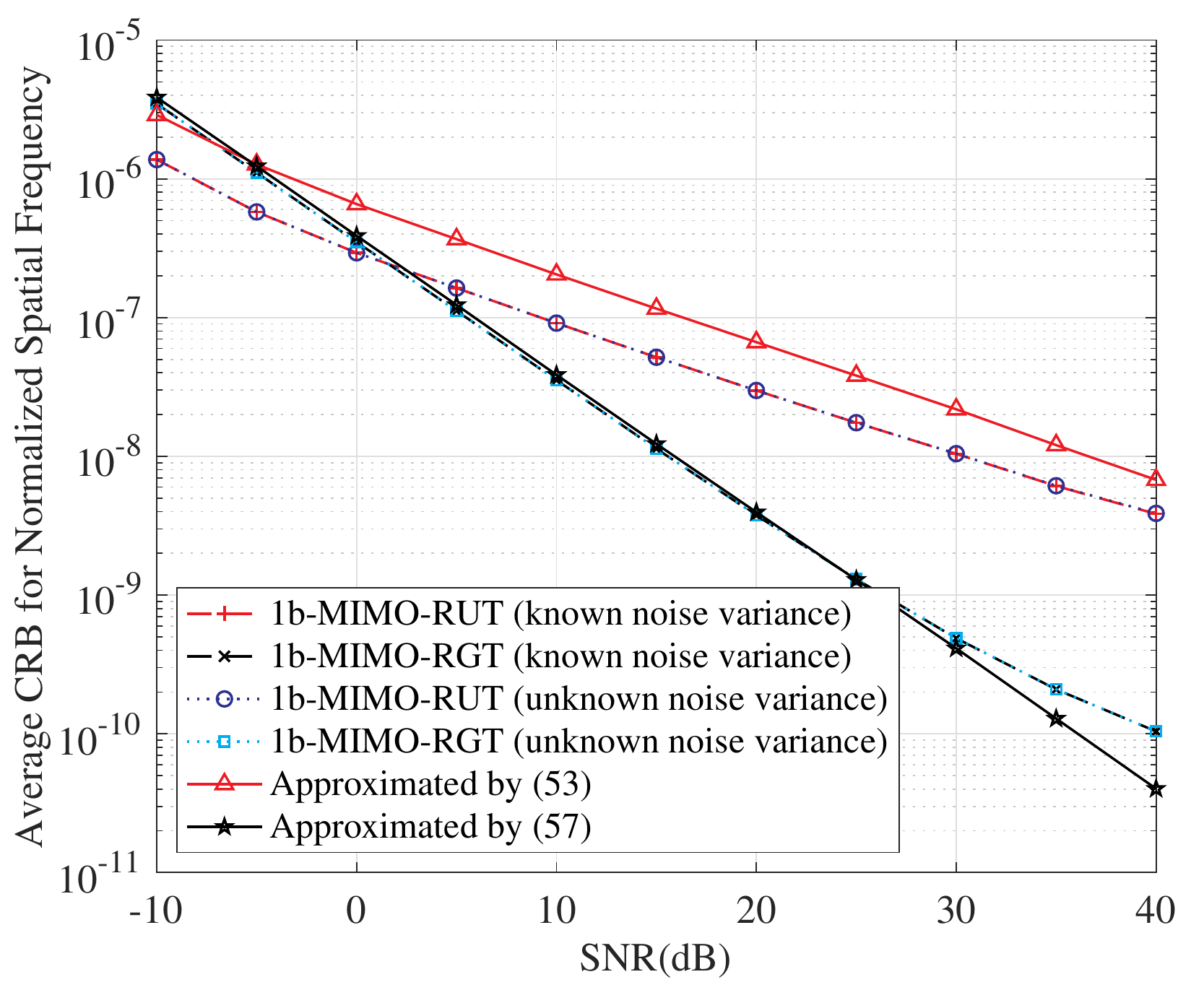}
\centering
\caption{A comparison of average CRB of NSF with known noise variance and that with unknown noise variance. The approximation formulations derived in Section VI.C are also shown.}
\end{figure}

\begin{figure}[!t]
\label{ddd}
\centering
\subfigure[]{
\includegraphics[width=1.8in]{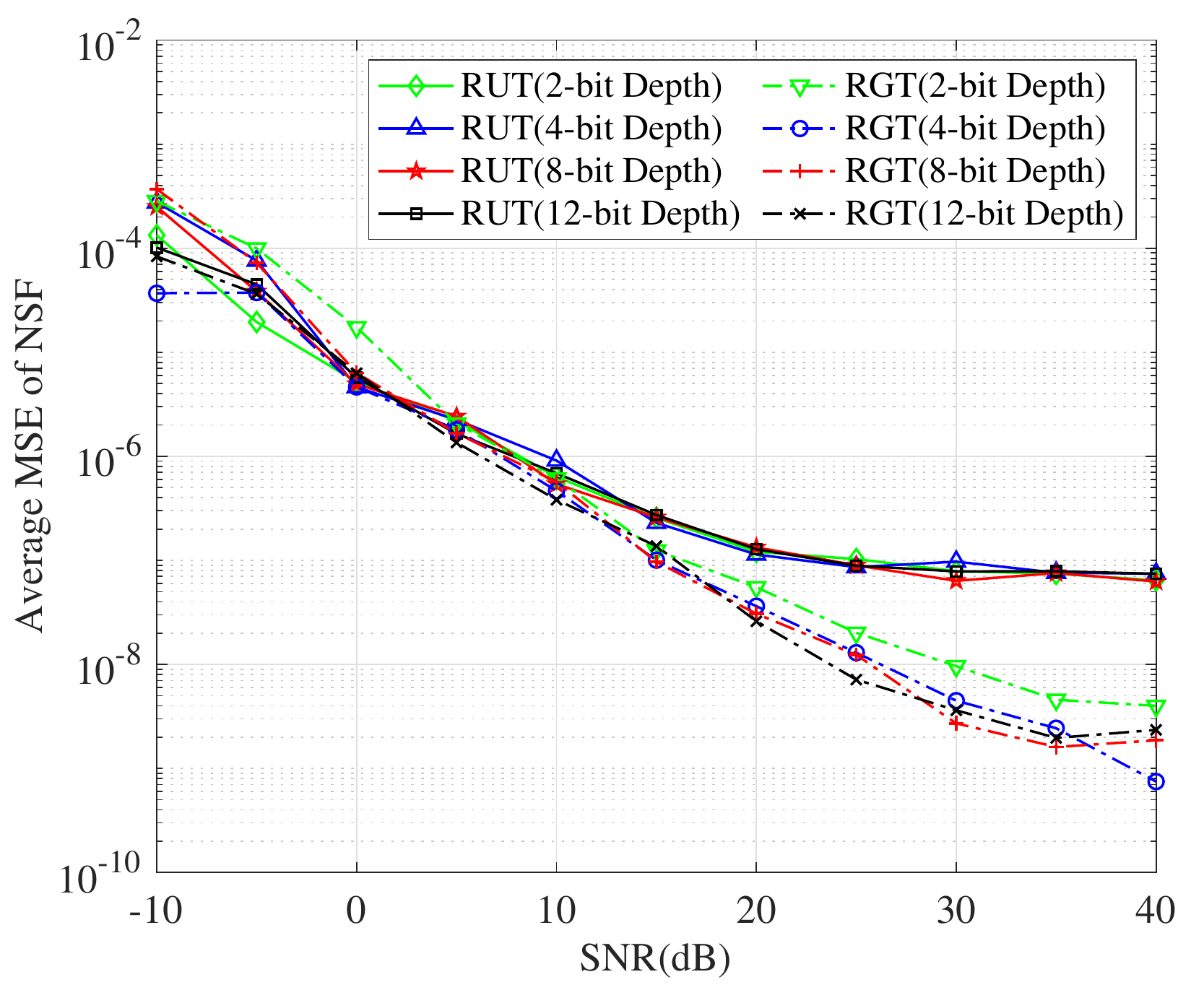}
}\\%
\subfigure[]{
\includegraphics[width=1.8in]{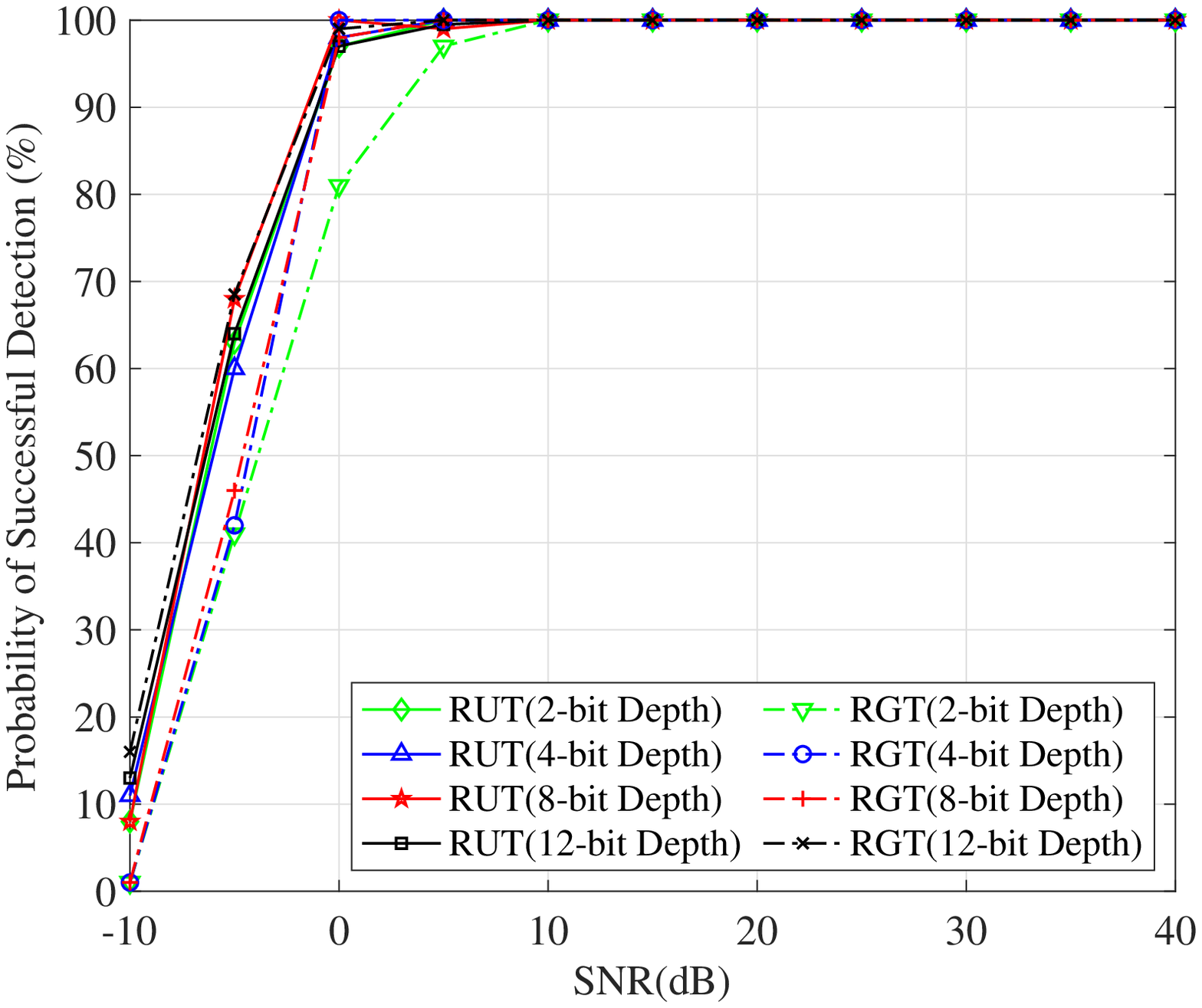}
}%
\centering
\caption{Estimation performance of the 1b-MIMO radar using thresholds with different bit depth. (a) Average MSE of NSF; (b) Probability of successful detection.}
\end{figure}
\begin{figure}[!t]
\label{aaa}
\centering
\subfigure[]{
\includegraphics[width=1.8in]{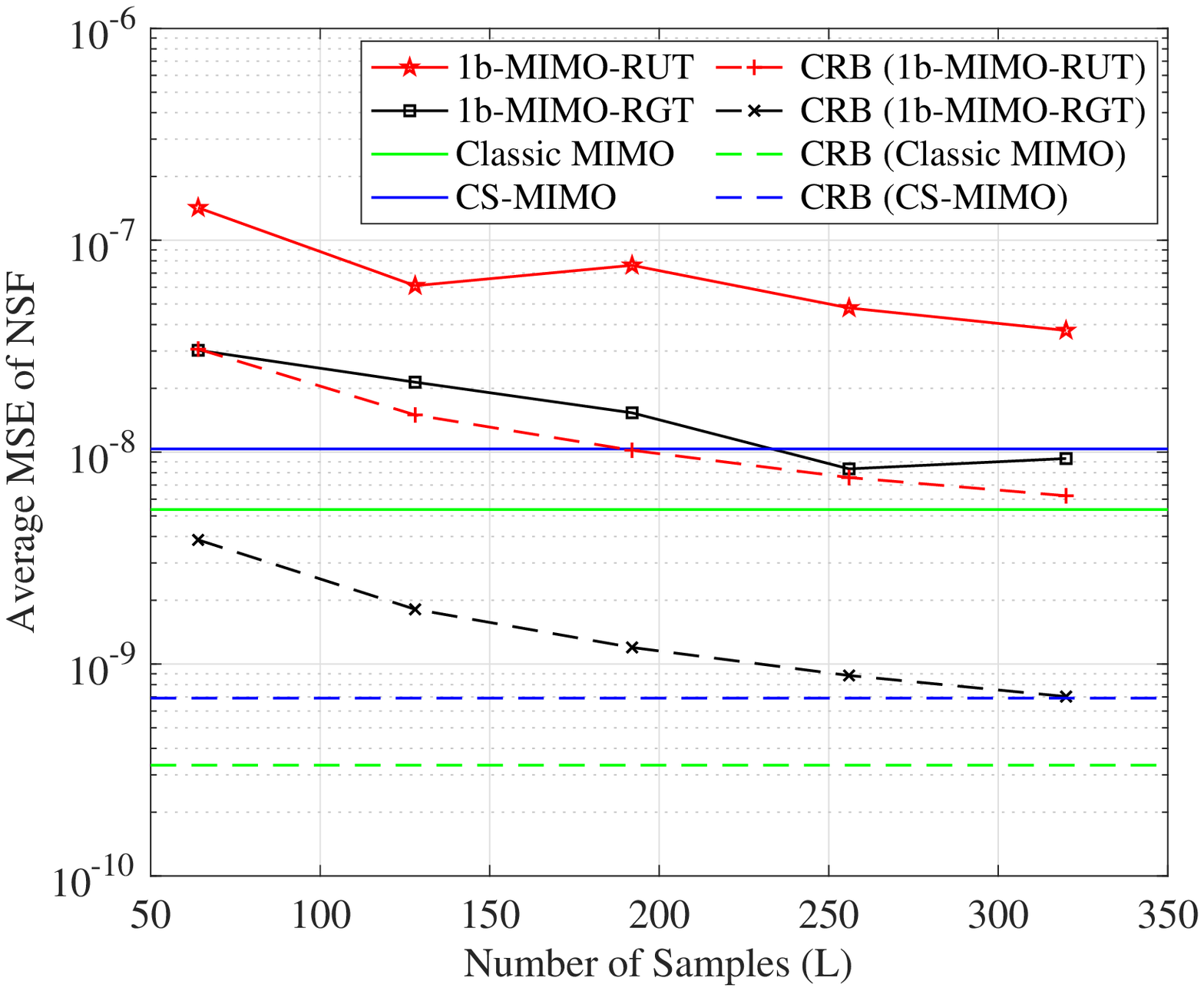}
}\\%
\subfigure[]{
\includegraphics[width=1.8in]{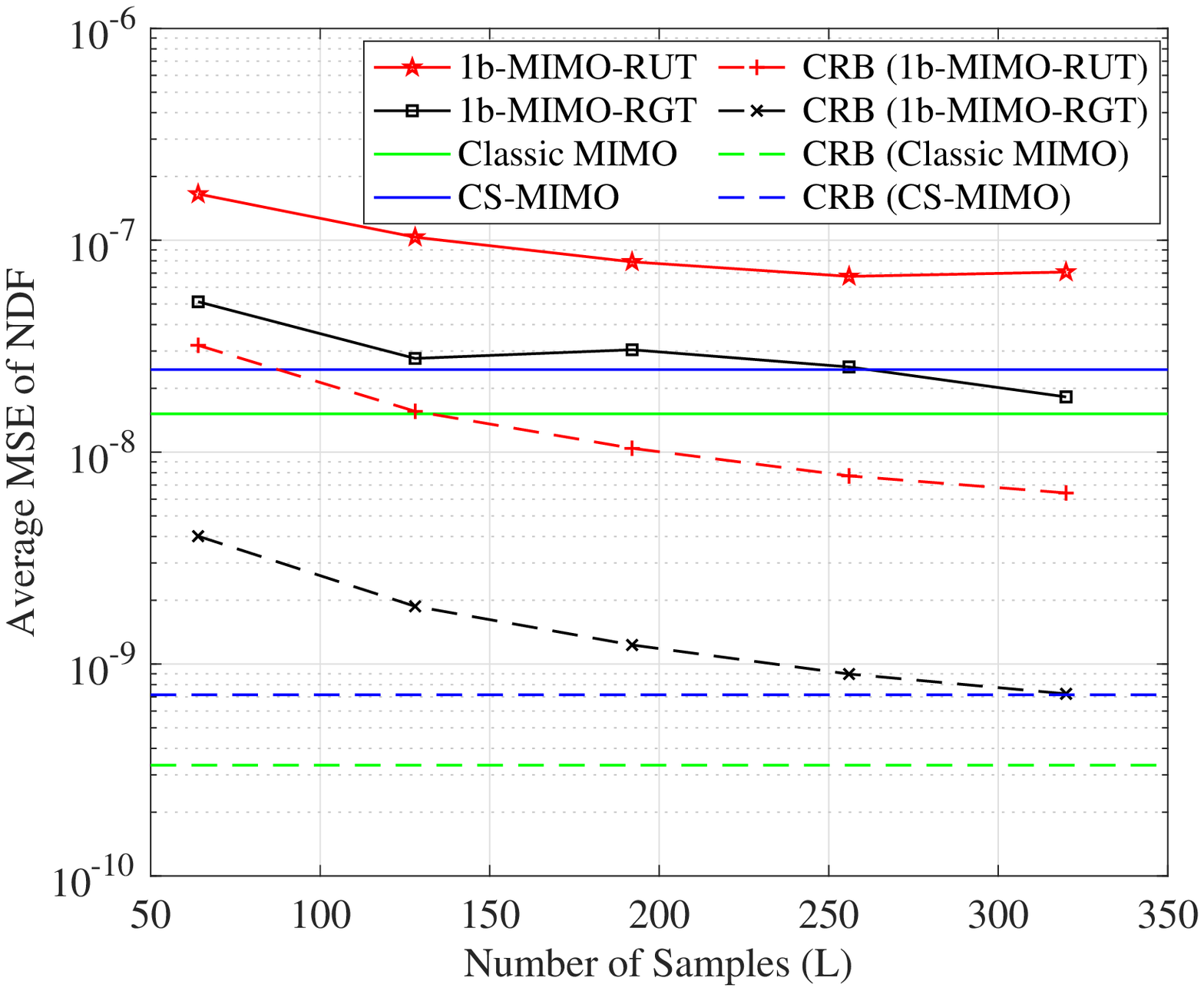}
}%

\centering
\caption{Average MSEs of NSF and NDF with respect to the number of samples $L$ when SNR=20dB. (a) Average MSE of NSF vs. L; (b) Average MSE of NDF vs. L.}
\end{figure}

\begin{figure}[!t]
\label{bbb}
\centering
\subfigure[]{
\includegraphics[width=1.8in]{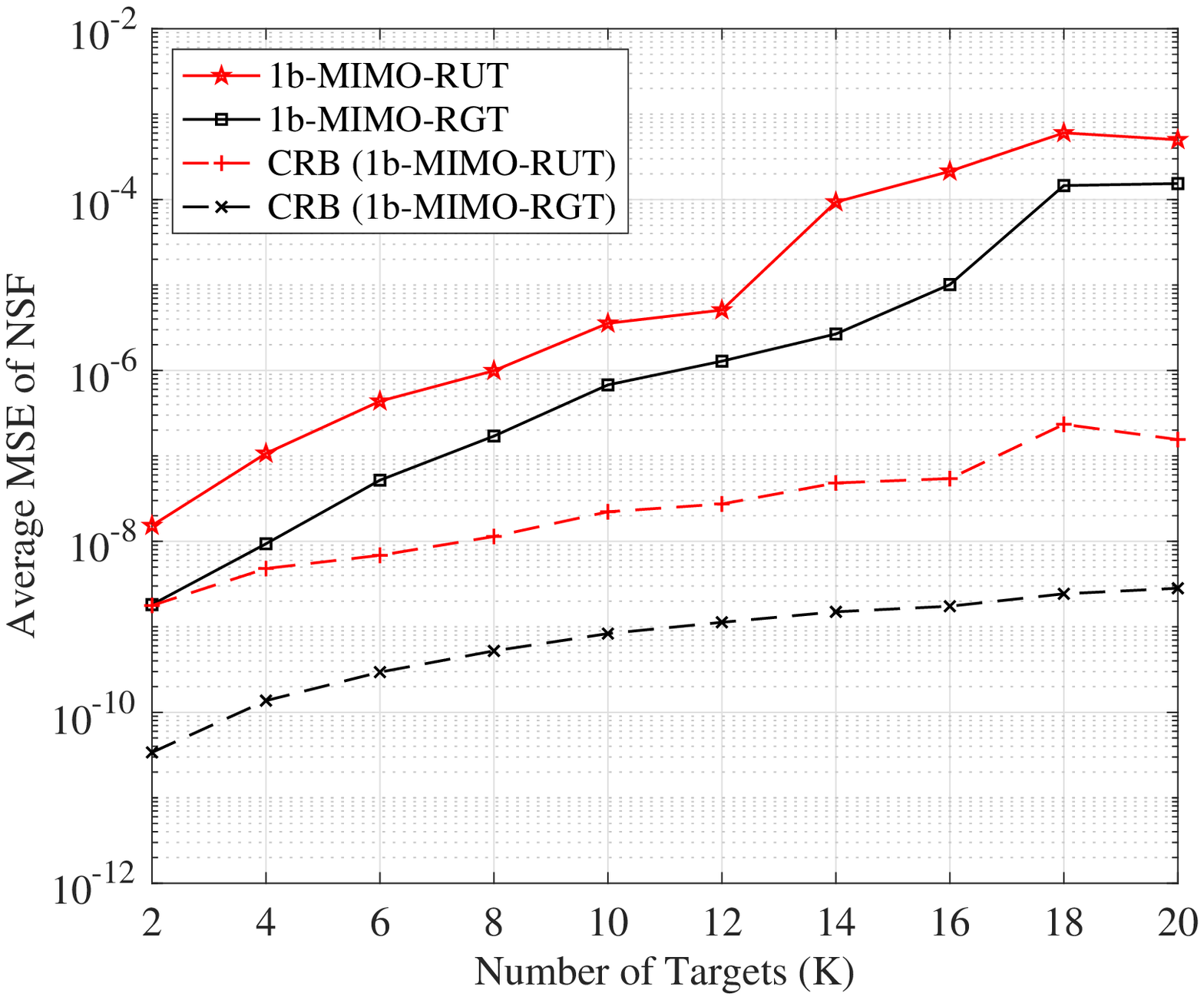}
}\\%
\subfigure[]{
\includegraphics[width=1.8in]{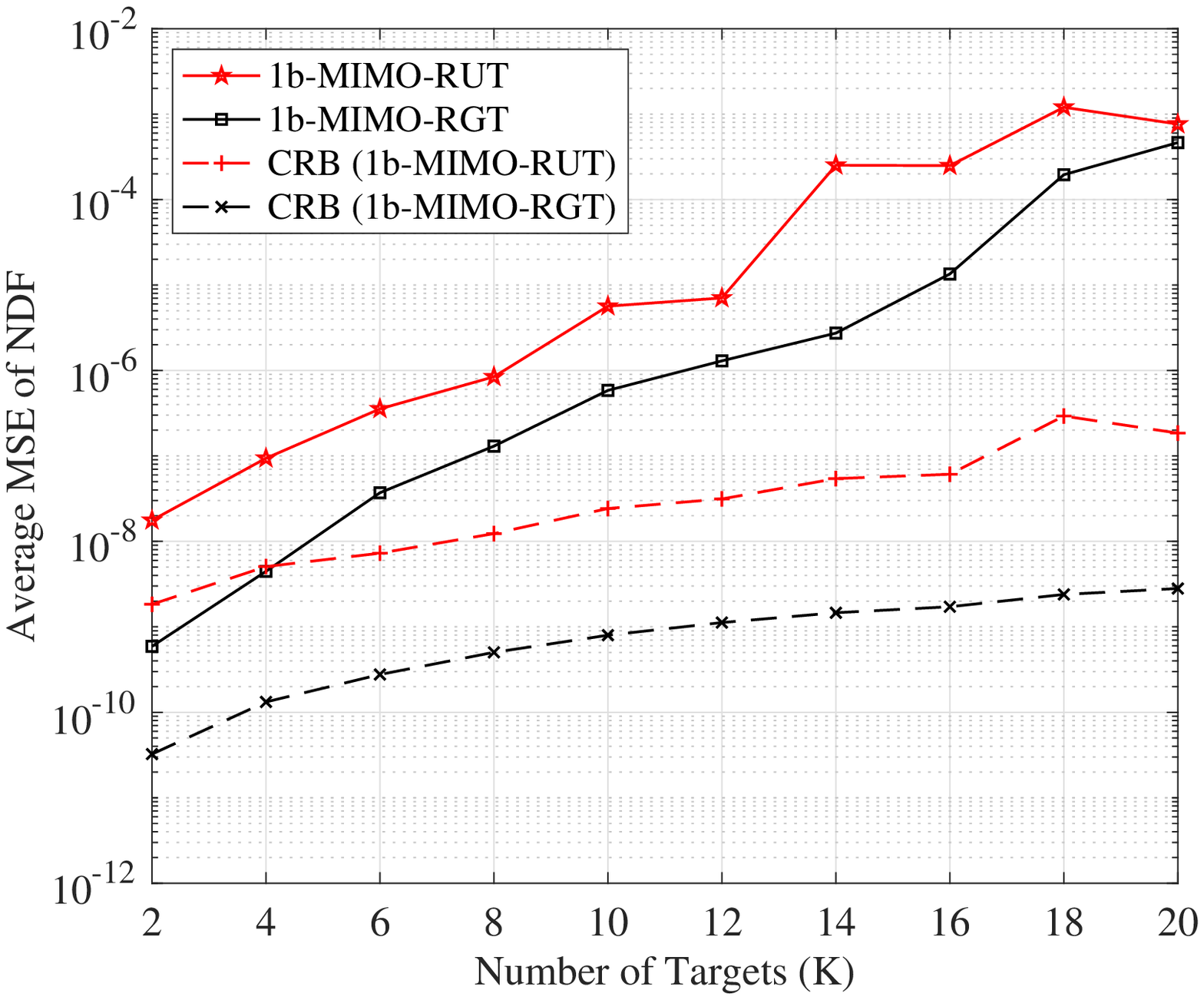}
}%
\centering
\caption{Average MSEs of NSF and NDF with respect to the number of targets $K$ when SNR=40dB. (a) Average MSE of NSF vs. K; (b) Average MSE of NDF vs. K.}
\end{figure}

The average MSEs of the magnitude and phase estimation are given in Fig. 3(d)$\sim$(e),
showing that the 1b-MIMO radar with time-varying thresholds can estimate the reflection coefficients of targets with high accuracy.
Specially, in the low SNR regime, the 1b-MIMO radar achieves even lower MSEs than its high-bit rivals.

In Fig. 3(f), we plot the normalized low-rank matrix recovery error, defined as $\|\mathbf{X}-\mathbf{\hat{X}}\|_F/\|\mathbf{X}\|_F$, for various SNR values.
As SNR increases, the recovery error in the one-bit cases remains close to the CS-MIMO and classic MIMO radars in the low-SNR regime, 
but it experiences severe performance degeneration in the high-SNR regime.
This plot also explains the performance gap between the 1b-MIMO radar and the high-bit quantized rivals in the high SNR regime.
Comparing the MSE and CRB curves between the RUT and RGT strategies in Fig. 3(a), (b), (d), and (e), 
it is clear that, by using the RGT strategy, the performance of the 1b-MIMO radar is significantly better in the high SNR regime.
Therefore, designing the optimal threshold to improve performance in the high SNR regime is an important issue for the 1b-MIMO radar.

In Fig. 4, the average CRB of NSF with known noise variance is compared with that with unknown noise variance. 
It is shown that the two CRB curves are almost the same. 
The CRBs of other parameters also have the similar results.
We omit them here due to length limitation of the paper.
We also show the CRB curves approximated by (53) and (57).
Although the approximated CRB given by (53) is a little higher than the true CRB of the RUT strategy, (57) approximates the CRB of the RGT strategy quite well.

In Fig. 5, we investigate the estimation performance by using the thresholds with different bit depth.
It is shown that, for the RGT strategy, we get lower MSE performance and higher PSD performance in the low SNR regime when using thresholds with higher bit depth.
For the RUT strategy, the 1b-MIMO radar using thresholds with different bit depth achieves almost the same MSE and PSD performance. 

We next increase the number of samples $L$ in the 1b-MIMO radar and demonstrate the MSE performance of the NSF and NDF with respect to $L$.
The code length of the corresponding quasi-orthogonal waveforms used in the 1b-MIMO radar also increases as the sampling rate increases.
Fig. 6 shows the MSEs of the NSF and NDF estimation, as well as their corresponding CRBs. 
The MSEs and CRBs of the classic MIMO radar and the CS-MIMO radar with fixed $L=64$ are also shown as benchmarks.
As $L$ increases, the 1b-MIMO radar improves its estimation performance.
Although the total number of bits in the 1b-MIMO radar is still lower than that of its 16-bit rivals, the performance of the 1b-MIMO-RGT radar is close to or even better than that of classic MIMO radar and CS-MIMO radar.
These results demonstrate that the performance of the 1b-MIMO radar can be significantly enhanced by increasing the sampling rate.
Moreover, it is worth pointing that the cost and energy consumption of a high-rate one-bit ADC may be lower than that of a low-rate high-bit ADC.
Therefore, it is possible for 1b-MIMO radar to achieve better performance than its high-bit rivals with lower cost and energy consumption.

Then, to illustrate the performance of the 1b-MIMO radar in resolving multiple closely-located targets, 
we consider a scenario where there are $K$ targets, located with the NSF and NDF pairs $\{\vartheta_0,\upsilon_0\}$, $\{\vartheta_0+1/MN,\upsilon_0+1/Q\}$, $\cdots$, $\{\vartheta_0+(K-1)/MN,\upsilon_0+(K-1)/Q\}$.
The magnitude of the targets' reflection coefficients are randomly distributed between $[0.2,1]$, and the phases of the reflection coefficients are randomly distributed between $[0,2\pi]$.
Thus the targets are separated by one resolution bin in the NSF-NDF plane.
We show the average MSEs of the NSF and NDF with respect to different $K$ when SNR=40dB in Fig. 7.
As the number of targets increases, the MSEs of the NSF and NDF estimation increase from $10^{-8}$ to $10^{-3}$, 
which means that the performance of the 1b-MIMO radar deteriorates as the number of targets increases.
The gaps between the achieved MSEs and the corresponding CRBs also become larger as $K$ increases. 
However, even when there are $K=12$ targets, 
the 1b-MIMO radar with both RUT and RGT can still achieve highly accurate NSF and NDF estimations with MSEs as low as $10^{-5}$.
When the number of targets exceeds $K=12$, the MSEs of the 1b-MIMO radar dramatically increase.
In practice, according to our previous definition of successful detection, the 1b-MIMO radar can achieve almost $100\%$ PSD when the number of targets does not exceed $K=12$.

\begin{table}
\caption{Different settings used in TABLE \ref{tab:2}.}
\label{tab:1}       
\begin{tabular}{llll} 
\hline\noalign{\smallskip}
Setting I & M=N=T=R=4& Q=16&L=64\\
Setting II & M=N=T=R=6& Q=36&L=64\\
Setting III & M=N=T=R=6&Q=36& L=128\\
Setting IV & M=N=T=R=8& Q=64&L=64\\
\noalign{\smallskip}\hline
\end{tabular}
\end{table}

\begin{table}
\caption{Running time of the different algorithms.}
\label{tab:2}       
\begin{tabular}{lllll} 
\hline\noalign{\smallskip}
\textbf{Settings} & Setting I & Setting II& Setting III&Setting IV\\
\noalign{\smallskip}\hline\noalign{\smallskip}
\textbf{ADMM}		&6.47s & 24.77s  & 37.28s   & 99.46s\\
\textbf{SDPT3}		& 40.58s  & -  & - & -\\
\textbf{MOSEK}		& 1.27s  & 16.24s  & 24.90s & 245.26s\\
\noalign{\smallskip}\hline
\end{tabular}
\end{table}

\begin{figure}[t]
\centering
\includegraphics[width=1.8in]{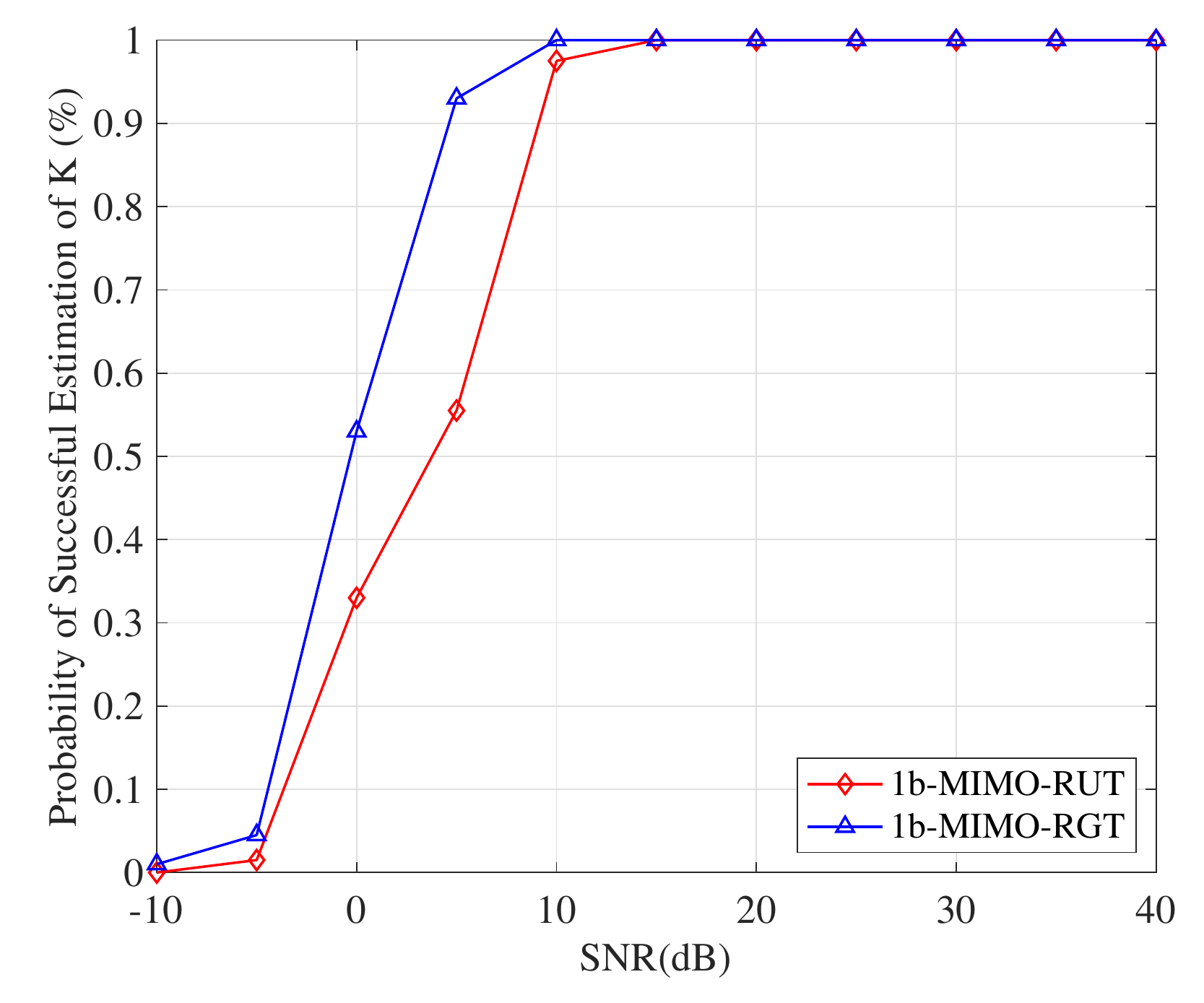}
\centering
\caption{Probability of successful estimation of $K$.}
\end{figure}
In Fig. 8, we exploit the distribution of the eigenvalues of $\mathbf{H}$ recovered in (\ref{eqn_RecX1}) to determine the number of targets $K$.
In our method, the estimate of $K$ is $\hat{K}$ satisfying $(\sum_{n=1}^{\hat{K}}\delta_n)/(\sum_{n=1}^{N}\delta_n) \geq \text{min}\{0.5+SNR/50, 0.9\}$, 
where $\delta_1,\cdots,\delta_{N}$ be the eigenvalues of $\mathbf{H}$ in descending order.
It is shown that the number of targets can be exactly estimated when $SNR>=10dB$.

Finally, we compare the running time of the proposed ADMM-based method with the "brute force" optimization using (\ref{eqn_RecX1}).
Two solvers, SDPT3 and MOSEK, in the CVX toolbox \cite{cvx} are used to solve the "brute force" optimization.
Table \ref{tab:1} lists the different settings of the 1b-MIMO radar used in this experiment.
The other experimental settings are similar to those in Fig. 3.
All experiments are carried out in Matlab R2018b on a Mac with a 2.9 GHz Intel Core i9 and 16 GB of RAM. 
The results are the average value of 200 individual trials.
It is seen from Table \ref{tab:2} that the proposed ADMM algorithm is computationally efficient and scales well as the problem dimension increases.
The "-" in SDPT3 represents that the algorithm is out of memory.
The commercial solver MOSEK performs well for Setting I, II, and III.
However, the running time of MOSEK increases significantly when the problem dimension becomes large.

\vspace{-0.4cm}
\section{Conclusion}
This paper discussed the possibility of employing one-bit sampling in MIMO radar to simplify the system's complexity as well as reduce its hardware cost and energy consumption.
To achieve high-resolution angle and Doppler frequency estimation, we developed the 1b-ANM-L1 method to account for the noise perturbations in one-bit sampling.
The core idea of the proposed method is that the effect of noise in one-bit sampling can be equivalent to that of sparse impulsive perturbation.
To accelerate the computation, an ADMM-based iterative algorithm was derived to compute the solution to the 1b-ANM-L1 problem.
The CRB performance of the 1b-MIMO radar with two different threshold strategies was analyzed, which showed that the RGT strategy can improve the performance in high SNR regime by utilizing \textit{a priori} information.
Simulation results showed that, while greatly reducing the amount of data, the 1b-MIMO radar can still achieve high-resolution parameter estimation.
The performance of the 1b-MIMO radar could be improved by slightly increasing the sampling rate.
Although the 1b-MIMO radar with RGT strategy can achieve performance close to its high-bit rivals, designing the optimal threshold strategy remains an open problem.

\appendices
\section{Proof of Equation (\ref{eqn_SDP})}

The proof of Equation (\ref{eqn_SDP}) relies on the classic Vandermonde decomposition of Toeplitz matrices \cite{Vander-1911,Stoica-2005}, which is summarized in the following lemma.

\setcounter{lem}{0}
\setcounter{equation}{0}
\renewcommand\thelem{A\arabic{lem}}
\renewcommand\theequation{A\arabic{equation}}

\begin{lem}\label{lemmaA1}
Suppose $\mathbf{T}\in\mathbb{C}^{L\times L}$ is a positive semidefinite (PSD) Toeplitz matrix, then $\mathbf{T}$ admits the following Vandermonde decomposition:
\begin{equation}
	\mathbf{T} = \mathbf{U}\mathbf{P}\mathbf{U}^H,
\end{equation}
where $\mathbf{P}\in\mathbb{C}^{R\times R}$ is positive definite diagonal matrix, $\mathbf{U} \in\mathbb{C}^{L\times R}$ is a Vandermonde matrix whose columns correspond to uniformly sampled complex sinusoids with different frequencies, and $R  =\rank(\mathbf{T})$. The decomposition is unique if $\mathbf{T}$ is rank-deficient, i.e., $R<L$.
\end{lem}

Now we give the proof of Equation (\ref{eqn_SDP}).

\begin{IEEEproof}
We first assume that $\|\mathbf{X}\|_{\mathcal{A},0}=K$, \emph{i.e.}, $\mathbf{X}$ admits an order-$K$ atomic decomposition $\mathbf{X}=\sum_{k=1}^{K}{a_{k}e^{j\phi_k}{\mathbf{w}(\varphi_{1k})}{\mathbf{v}^H(\varphi_{2k})}}$, where the pairs of parameters $\{\varphi_{11},\varphi_{21}\}$, $\cdots$, $\{\varphi_{1K},\varphi_{2K}\}$ are different.
Then we can construct the following matrix satisfying the constraint in (\ref{eqn_SDP}):
\begin{equation}
\begin{split}
\mathbf{H}& =\begin{bmatrix}
\sum_{k=1}^{K}{a_{k}{\mathbf{w}(\varphi_{1k})}{\mathbf{w}^H(\varphi_{1k})}} & \mathbf{X}\\
\mathbf{X}^H & \sum_{k=1}^{K}{a_{k}{\mathbf{v}(\varphi_{2k})}{\mathbf{v}^H(\varphi_{2k})}}\\
\end{bmatrix}\\
&=\sum_{k=1}^{K}a_{k}\begin{bmatrix}
e^{j\phi_k}{\mathbf{w}(\varphi_{1k})}\\
\mathbf{v}(\varphi_{2k})\end{bmatrix}
\begin{bmatrix}
e^{j\phi_k}{\mathbf{w}(\varphi_{1k})}\\
\mathbf{v}(\varphi_{2k})\end{bmatrix}^H\succeq0.
\end{split}
\end{equation}
It follows that $K^\ast\leq\textrm{rank}(\mathbf{H})=K=\|\mathbf{X}\|_{\mathcal{A},0}$, where $K^\ast$ is the optimal solution of the right hand side (RHS) of (\ref{eqn_SDP}).

On the other hand, when the RHS of (\ref{eqn_SDP}) achieves the optimal value $K^\ast$ at the optimizer $(\mathbf{u}_1^\ast,\mathbf{u}_2^\ast)$, there exists a rank-$K^\ast$ matrix $\mathbf{H}^\ast$:
\begin{equation}
\mathbf{H}^\ast=\begin{bmatrix}
\mathcal{T}(\mathbf{u}_1^\ast) & \mathbf{X}\\
\mathbf{X}^H & \mathcal{T}(\mathbf{u}_2^\ast)
\end{bmatrix},
\end{equation}
\emph{i.e.}, $\mathbf{H}^\ast$ can be decomposed as:
\begin{equation}
\mathbf{H}^\ast=\begin{bmatrix}
\mathbf{H}_1\\
\mathbf{H}_2\\
\end{bmatrix}\begin{bmatrix}
\mathbf{H}_1\\
\mathbf{H}_2\\
\end{bmatrix}^H,
\end{equation}
where $\mathbf{H}_1 \in\mathbb{C}^{MN\times K^\ast}$ and $\mathbf{H}_2 \in\mathbb{C}^{Q\times K^\ast}$.

According to Lemma \ref{lemmaA1}, the Toeplitz matrices $\mathcal{T}(\mathbf{u}_1^\ast)$ and $\mathcal{T}(\mathbf{u}_2^\ast)$ admit the following Vandermonde decompositions:
\begin{equation}
\mathcal{T}(\mathbf{u}_1^\ast)=\mathbf{U}_1\mathbf{P}_1\mathbf{U}_1^H,
\end{equation}
\begin{equation}
\mathcal{T}(\mathbf{u}_2^\ast)=\mathbf{U}_2\mathbf{P}_2\mathbf{U}_2^H,
\end{equation}
where $\mathbf{U}_1=[\mathbf{w}(\varphi_{11}^\ast),\mathbf{w}(\varphi_{12}^\ast),\cdots,\mathbf{w}(\varphi_{1K_1}^\ast)]$ and $\mathbf{U}_2=[\mathbf{v}(\varphi_{21}^\ast),\mathbf{v}(\varphi_{22}^\ast),\cdots,\mathbf{v}(\varphi_{2K_2}^\ast)]$ with $K_1,K_2\leq K^\ast$, and $\mathbf{P}_1\in\mathbb{C}^{K_1\times K_1}$ and $\mathbf{P}_2\in\mathbb{C}^{K_2\times K_2}$ are two positive definite diagonal matrices.

Since $\mathcal{T}(\mathbf{u}_1^\ast)=\mathbf{H}_1\mathbf{H}_1^H$ and $\mathcal{T}(\mathbf{u}_2^\ast)=\mathbf{H}_2\mathbf{H}_2^H$, it holds that $\mathbf{H}_1=\mathbf{U}_1\mathbf{P}_1^{1/2}\mathbf{O}_1$, $\mathbf{H}_2=\mathbf{U}_2\mathbf{P}_2^{1/2}\mathbf{O}_2$, where $\mathbf{O}_1 \in \mathbb{C}^{K_1\times K^\ast}$, $\mathbf{O}_2 \in \mathbb{C}^{K_2\times K^\ast}$ and $\mathbf{O}_1\mathbf{O}_1^H=\mathbf{I}_{K_1}$, $\mathbf{O}_2\mathbf{O}_2^H=\mathbf{I}_{K_2}$. 
Then $\mathbf{X}$ can be represented as
\begin{equation}
\label{eqn_Xdecomp}
\mathbf{X}=\mathbf{U}_1\mathbf{P}_1^{1/2}\mathbf{O}_1\mathbf{O}_2^H\mathbf{P}_2^{1/2}\mathbf{U}_2^H.
\end{equation}
This means that each column of $\mathbf{X}$ lies in the range space spanned by $\{\mathbf{w}(\varphi_{1k}^\ast)\}_{k=1}^{K_1}$.
Similarly, each column of $\mathbf{X}^H$ lies in the range space spanned by $\{\mathbf{v}(\varphi_{2k}^\ast)\}_{k=1}^{K_2}$.

Since $\mathbf{X}$ has the atomic representation defined in (\ref{eqn_Atoml0norm}), we assume that $\mathbf{X}=\sum_{k=1}^K {\beta_k}\mathbf{w}(\varphi_{1k})\mathbf{v}(\varphi_{2k})$, where $\beta_k=a_{k}e^{j\phi_k}$.
According to (\ref{eqn_Xdecomp}), it holds that $\varphi_{1k}$ and $\varphi_{2k}$ must belong to $\{\varphi_{11}^\ast,\varphi_{12}^\ast,\cdots,\varphi_{1K_1}^\ast\}$ and $\{\varphi_{21}^\ast,\varphi_{22}^\ast,\cdots,\varphi_{2K_2}^\ast\}$, respectively. 
Therefore, $\mathbf{X}$ can also be expressed as
\begin{equation}
\label{eqn_Xdecomp2}
\mathbf{X}=\tilde{\mathbf{U}}_1\mathbf{B}\tilde{\mathbf{U}}_2^H=\mathbf{U}_1\mathbf{Z}_1\mathbf{B}\mathbf{Z}_2^H\mathbf{U}_2^H,
\end{equation}
where $\mathbf{B}=\diag([\beta_1,\cdots,\beta_K])$, and $\tilde{\mathbf{U}}_1\in\mathbb{C}^{MN\times K}$ and $\tilde{\mathbf{U}}_2\in\mathbb{C}^{Q\times K}$ are matrices with rearranged columns (and maybe replicated columns) of $\mathbf{U}_1$ and $\mathbf{U}_2$, respectively.
$\mathbf{Z}_1\in\mathbb{R}^{K_1\times K}$ and $\mathbf{Z}_2\in \mathbb{R}^{K_2\times K}$ are two matrices representing the rearrangement and replication operations. 
Comparing (\ref{eqn_Xdecomp}) with (\ref{eqn_Xdecomp2}), it is equivalent to require that
\begin{equation}
\label{eqn_Xdecomp3}
\mathbf{Z}_1\mathbf{B}\mathbf{Z}_2^H=\mathbf{P}_1^{1/2}\mathbf{O}_1\mathbf{O}_2^H\mathbf{P}_2^{1/2}.
\end{equation}
If we let $\mathbf{Z}_1\mathbf{B}^{1/2} = \mathbf{P}_1^{1/2}\mathbf{O}_1$ and $\mathbf{Z}_2\mathbf{B}^{1/2} = \mathbf{P}_2^{1/2}\mathbf{O}_2$,
(\ref{eqn_Xdecomp3}) is satisfied.
In this case, $K = K^\ast$.
According to the definition of $\|\mathbf{X}\|_{\mathcal{A},0}$, we have
$\|\mathbf{X}\|_{\mathcal{A},0}\leq K = K^\ast$.

Therefore, we prove that $\|\mathbf{X}\|_{\mathcal{A},0} = K^\ast$,
where $K^\ast$ is the optimal solution of the RHS of (\ref{eqn_SDP}).
	
\end{IEEEproof}

\vspace{-0.3cm}
\section{CRB for 1-bit Quantized Data with Unknown Noise Variance}
\setcounter{lem}{0}
\setcounter{equation}{0}
\renewcommand\thelem{B\arabic{lem}}
\renewcommand\theequation{B\arabic{equation}}

Let $L$ be the log-likelihood of the 1-bit quantized data $\{\mathbf{z}_q\}_{q\in\Omega^p}$.
According to (\ref{eqn_PMF}), $L$ is given as
\begin{equation}
\label{eqn_loglike_1b}
L = L_r + L_i,
\end{equation}
where
\begin{equation}
\label{eqn_loglike_1br}
\begin{split}
	L_r &= \sum_{q\in\Omega^p}\sum_{n=1}^{LR}[\frac{1+\Re\{[\mathbf{z}_q]_n\}}{2}\log\Phi\left(\frac{[\mathbf{r}_q-\mathbf{h}^r_q]_n}{\sigma}\right)\\
	&+\frac{1-\Re\{[\mathbf{z}_q]_n\}}{2}\log\left(1-\Phi\left(\frac{[\mathbf{r}_q-\mathbf{h}^r_q]_n}{\sigma}\right)\right)],
\end{split}
\end{equation}
and $L_i$ is expressed similarly by replacing $\Re\{[\mathbf{z}_q]_n\}$ and $\mathbf{r}_q-\mathbf{h}^r_q$ with $\Im\{[\mathbf{z}_q]_n\}$ and $\mathbf{i}_q-\mathbf{h}^i_q$, respectively.

Then the derivative of $L_r$ with respective to $\sigma$ is given as
\begin{equation}
	\frac{\partial L_r}{\partial \sigma} =  \sum_{q\in\Omega^p}\sum_{n=1}^{LR} \left[\frac{1+\Re\{[\mathbf{z}_q]_n\}}{2\Phi_{q,n}^r}
	-\frac{1-\Re\{[\mathbf{z}_q]_n\}}{2(1-\Phi_{q,n}^r)}\right]\frac{\partial\Phi_{q,n}^r}{\partial\sigma},
\end{equation}
where $\Phi_{q,n}^r=\Phi\left(\frac{[\mathbf{r}_q-\mathbf{h}^r_q]_n}{\sigma}\right)$.
Similar result can be derived for $\frac{\partial L_i}{\partial \sigma}$.
Therefore, we can calculate the following result
\begin{equation}\label{Eqn_Esigma_1b}
\begin{split}
  \mathbb{E}\left[\left(\frac{\partial{L}}{\partial\sigma}\right)^2\right] &= \mathbb{E}\left[\left(\frac{\partial{L_r}}{\partial\sigma}\right)^2\right] +	\mathbb{E}\left[\left(\frac{\partial{L_i}}{\partial\sigma}\right)^2\right],\\
\end{split}
\end{equation}
where
\begin{equation}
\begin{split}
	&\mathbb{E}\left[\left(\frac{\partial{L_r}}{\partial\sigma}\right)^2\right] \\
	=& \sum_{q\in\Omega^p}\sum_{n=1}^{LR} \mathbb{E}\left[\frac{1+\Re\{[\mathbf{z}_q]_n\}}{2}\right]^2\left(\frac{1}{\Phi_{q,n}^r}\frac{\partial\Phi_{q,n}^r}{\partial\sigma}\right)^2\\
	 &+ \mathbb{E}\left[\frac{1-\Re\{[\mathbf{z}_q]_n\}}{2}\right]^2\left(\frac{1}{1-\Phi_{q,n}^r}\frac{\partial\Phi_{q,n}^r}{\partial\sigma}\right)^2\\
	 =&\sum_{q\in\Omega^p}\sum_{n=1}^{LR}\left(\frac{1}{\Phi_{q,n}^r}+\frac{1}{1-\Phi_{q,n}^r}\right)\left(\frac{\partial\Phi_{q,n}^r}{\partial\sigma}\right)^2\\
	 =&\frac{2}{\sigma^2}\sum_{q\in\Omega^p}\sum_{n=1}^{LR}\omega\left(\frac{[\mathbf{r}_q-\mathbf{h}^r_q]_n}{\sigma}\right)\frac{[\mathbf{r}_q-\mathbf{h}^r_q]_n^2}{\sigma^2},
\end{split}
\end{equation}
\begin{equation}
\begin{split}
	&\mathbb{E}\left[\left(\frac{\partial{L_i}}{\partial\sigma}\right)^2\right] \\
	=& \frac{2}{\sigma^2}\sum_{q\in\Omega^p}\sum_{n=1}^{LR}\omega\left(\frac{[\mathbf{i}_q-\mathbf{h}^i_q]_n}{\sigma}\right)\frac{[\mathbf{i}_q-\mathbf{h}^i_q]_n^2}{\sigma^2}.
\end{split}
\end{equation}
In the above derivation, we apply the result that $\mathbb{E}\left[\frac{\partial L_r}{\partial \sigma}\frac{\partial L_i}{\partial \sigma}\right] =0$.

Furthermore, since $\mathbb{E}\left[\frac{\partial L_r}{\partial \sigma}\frac{\partial L_i}{\partial \theta_i}\right]=\mathbb{E}\left[\frac{\partial L_r}{\partial \theta_i}\frac{\partial L_i}{\partial \sigma}\right]=0$, we can derive that
\begin{equation}\label{Eqn_Esigmatheta_1b}
\begin{split}
	\mathbb{E}\left[\frac{\partial L}{\partial \sigma}\frac{\partial L}{\partial \theta_i}\right] &= \mathbb{E}\left[\frac{\partial{L_r}}{\partial\sigma}\frac{\partial{L_r}}{\partial\theta_i}\right] +	\mathbb{E}\left[\frac{\partial{L_i}}{\partial\sigma}\frac{\partial{L_i}}{\partial\theta_i}\right],\\
\end{split}
\end{equation}
where
\begin{equation}
	\begin{split}
		&\mathbb{E}\left[\frac{\partial{L_r}}{\partial\sigma}\frac{\partial{L_r}}{\partial\theta_i}\right]\\
				=&-\frac{2}{\sigma^2}\sum_{q\in\Omega^p}\sum_{n=1}^{LR}\omega\left(\frac{[\mathbf{r}_q-\mathbf{h}^r_q]_n}{\sigma}\right)\frac{[\mathbf{r}_q-\mathbf{h}^r_q]_n}{\sigma}\frac{\partial[\mathbf{r}_q]_n}{\partial\theta_i},
	\end{split}
\end{equation}
\begin{equation}
\begin{split}
		&\mathbb{E}\left[\frac{\partial{L_i}}{\partial\sigma}\frac{\partial{L_i}}{\partial\theta_i}\right]\\
		 =& -\frac{2}{\sigma^2}\sum_{q\in\Omega^p}\sum_{n=1}^{LR}\omega\left(\frac{[\mathbf{i}_q-\mathbf{h}^i_q]_n}{\sigma}\right)\frac{[\mathbf{i}_q-\mathbf{h}^i_q]_n}{\sigma}\frac{\partial[\mathbf{i}_q]_n}{\partial\theta_i}.
\end{split}
\end{equation}

Let $\mathbf{I}(\sigma,\boldsymbol{\theta}) = \left[\mathbb{E}\left[\frac{\partial L}{\partial \sigma}\frac{\partial L}{\partial \theta_1}\right],\cdots,\mathbb{E}\left[\frac{\partial L}{\partial \sigma}\frac{\partial L}{\partial \theta_{4K}}\right]\right]^T$ and $\mathbf{I}(\sigma) =\mathbb{E}\left[\left(\frac{\partial{L}}{\partial\sigma}\right)^2\right]$,
where $\theta_1, \cdots, \theta_{4K}\in\boldsymbol{\theta}$.

By applying (\ref{Eqn_Esigma_1b}) and (\ref{Eqn_Esigmatheta_1b}), we can derive the Fisher information matrix given in (\ref{Eqn_FIM_unq_noise_1b}).

\balance
\bibliographystyle{IEEEtran}
\bibliography{IEEEabrv,MIMO,Onebit}

\end{document}